\documentclass[12pt, a4paper]{article}
\usepackage{amsmath, hyperref,  anysize, nameref, bm, graphicx, setspace}
\newtheorem{assumption}{Assumption}
\newtheorem{theorem}{Theorem}

\newtheorem{lemma}{Lemma}
\title{Estimation methods for elementary chirp model parameters}
\author{Anjali~Mittal, ~Rhythm~Grover, ~Debasis~Kundu, and~Amit~Mitra}
\date{}

\begin{document}
	\maketitle
	\begin{abstract}
		In this paper, we propose some estimation techniques to estimate the elementary chirp model parameters, which are encountered in sonar, radar, acoustics, and other areas. We derive asymptotic theoretical properties of least squares estimators and approximate least squares estimators for the one-component elementary chirp model. It is proved that the proposed estimators are strongly consistent and follow the normal distribution asymptotically. We also suggest how to obtain proper initial values for these methods. The problem of finding initial values is a difficult problem when the number of components in the model is large, or when the signal-to-noise ratio is low, or when two frequency rates are close to each other. We propose sequential procedures to estimate the multiple-component elementary chirp model parameters. We prove that the theoretical properties of sequential least squares estimators and sequential approximate least squares estimators coincide with those of least squares estimators and approximate least squares estimators, respectively. To evaluate the performance of the proposed estimators, numerical experiments are performed. It is observed that the proposed sequential estimators perform well even in situations where least squares estimators do not perform well. We illustrate the performance of the proposed sequential algorithm on a bat data.
	\end{abstract}
	
	\noindent \textit{\textbf{Index Terms-}}	Chirp model, approximate least squares, least squares, sequential least squares, frequency rate, consistency, asymptotic normality.
	
	\section{Introduction}
	We consider the following model:
	\begin{equation} \label{multicompmodel}
		y\left( t\right) = \sum_{k=1}^{p} A_{k}^{0} e^{i\beta_{k}^{0} t^{2}}+ \epsilon\left( t\right); ~ t=1,\dots, N ,
	\end{equation}		
	where, \(A_{k}^{0}\)s are the complex-valued non-zero amplitude parameters and \( i=\sqrt{-1} \). The \( \beta_{k}^{0} \)s are the frequency rate parameters, which strictly lie between \( 0 \) and \( 2\pi \) and they are distinct. Also, \( \epsilon\left( t\right)^\prime \)s  are the complex-valued noise random variables present in the observed signal \(  y\left(t \right) \). Detailed assumptions on \( \epsilon\left( t\right)^\prime \)s  are stated later (see assumption \ref{ass1}).
	
	\noindent Further, $p$ is the number of chirp components and is assumed to be known. For the observed data $y\left( 1\right), y\left( 2\right), \dots, y\left(N \right)$, the problem here is to estimate the unknown amplitude parameters $A_{k}^{0}$s and the unknown frequency rates $\beta_{k}^{0}$s. 
	
	\noindent The model (\ref{multicompmodel}) is named as multiple-component elementary chirp model. Although the model \eqref{multicompmodel} is widely applicable, the literature on this model is rather limited (see, for example, Mboup and Adali \cite{mboup} and Casazza and Fickus \cite{fickus} for few notable exceptions). Here, estimation of the chirp rate i.e. the frequency rate is of utmost importance. Some of the applications where chirp rate estimation is considered as of prime importance can be found in sonar pulse detection \cite{crappsonar}, in  micro-Doppler signal analysis \cite{crappmicro}, in acoustic signal analysis \cite{crappacoustic}, in focusing on the synthetic aperture radar images \cite{crappradar}, and many more. There are some estimation methods which mainly aim on the estimation of the instantaneous frequency rate (IFR), which is twice the chirp rate. Estimator based on cubic phase function (CPF) \cite{ cpf2} is one of them and other methods motivated by CPF such as Nonparametric Chirp-Rate estimator based on CPF \cite{crestnpi}, viterbi algorithm \cite{crestva}, integrated CPF (ICPF) \cite{icpf} and product CPF (PCPF) \cite{pcpf}, to name a few, have been discussed in the literature.\\
	
	\noindent In this paper, we propose some estimation methods to estimate the elementary chirp model \eqref{multicompmodel} parameters. We propose least squares estimators (LSEs), approximate least squares estimators (ALSEs), sequential LSEs and sequential ALSEs, study their theoretical asymptotic properties and compare their numerical performances. Model \eqref{multicompmodel} is a non-linear regression model, as should be noted. In the literature, theoretical results on the general non-linear regression model have been established by Jennrich \cite{jennrich} and Wu \cite{wu}. It has been observed that the sufficient conditions of Jennrich \cite{jennrich} and Wu \cite{wu} are not satisfied by model \eqref{multicompmodel} for the LSEs to be consistent. Thus, one cannot apply the results of Jennrich \cite{jennrich} and Wu \cite{wu} directly to establish the theoretical properties of the LSEs.\\
	
	\noindent It is established that the least squares estimation method and the approximate least squares estimation method provide the same optimal rates of convergence for the amplitude parameters and the frequency rate parameters, that is $O_{p}\left( N^{-\frac{1}{2}} \right) $ and  $O_{p}\left( N^{-\frac{5}{2}} \right) $, respectively. Here, $O_{p}\left( N^{-\delta} \right) $ means $N^{\delta} O_{p}\left( N^{-\delta} \right) $ is bounded in probability. In all the proposed estimation methods, we have to perform non-linear optimization for which we need to employ a numerical method. To employ a numerical technique, a set of initial values of the non-linear parameters is required. In the absence of good initial values (near to the true parameter values), due to high non-linearity of the least squares surface, the algorithm rather than converging to a global minimum, may converge to a local minimum, see Rice and Rosenblatt \cite{ricerosenblatt}. Therefore, to choose good initial values, we use the conventional grid search method. For the model \eqref{multicompmodel}, it is observed that the general-purpose iterative procedures like Newton-Raphson, Gauss-Newton, or their different versions, require a long time to converge to the LSEs even from a set of good initial values. Therefore, we use downhill simplex method, to compute the LSEs and ALSEs, efficiently. \\
	
	\noindent To obtain the initial values for the proposed methods in case of model \eqref{multicompmodel}, we have to do a multi-dimensional grid search which is a numerically intense problem in itself. Further, when the variance of the error random variable is high or when the two frequency rates are close to each other, multi-dimensional grid search may result in the initial values which are not close to the true parameter values. This may lead to incorrect parameter estimates. To overcome this problem, we propose a sequential least squares estimation method and a sequential approximate least squares estimation method. These sequential methods lower the computational complexity by reducing the $p$-dimensional optimization problem to $p$, 1-D optimization problems.  We also establish the theoretical properties of sequential LSEs and sequential ALSEs and find that they have the same theoretical properties as their respective LSEs and ALSEs.\\
	
	\noindent Furthermore, we perform extensive simulation studies to evaluate the performance of the proposed estimation methods for various sample sizes and error variances. We also obtain frequency rate estimates using other standard estimation methodologies and assess the effectiveness of the proposed methods in comparison to these techniques. For the one-component elementary chirp model, dechirping method \cite{dechirp} and CPF method \cite{cpf2} and for the multiple-component elementary chirp model, dechirping method and PCPF method \cite{pcpf} have been used for the comparative study. It is noted that the proposed estimators perform quite satisfactorily. The mean squared errors (MSEs) of the proposed estimators are close to their respective theoretical asymptotic variances. Another interesting observation that came out of these experiments is that the proposed sequential estimators are able to resolve the frequency rates even when two frequency rates are close to each other whereas LSEs are unable to do so at times. This motivates us to use the proposed sequential estimators for the implementation purposes due to their good theoretical properties and also excellent simulation results in different presented scenarios. We also show how well the proposed sequential estimators work by fitting the elementary chirp model to a real-world data set.\\
	
	\noindent The rest of the paper is structured as follows. In the next section, we present the statistical properties of the LSEs and the ALSEs for the one-component elementary chirp model. In section \ref{multicompsec}, we present theoretical properties of the sequential LSEs and the sequential ALSEs for the multiple-component elementary chirp model \eqref{multicompmodel}. We provide simulation results to validate the theoretical results of the proposed methods in section \ref{simdatasec}. Real data analysis is presented in section \ref{realdatana}. In section \ref{concludesec}, the paper is concluded. All the necessary proofs and results are presented in the appendices section of supplementary material.
	
	\section{One-Component Elementary Chirp Model} \label{onecompsec}
	\noindent In this section, we consider the following one-component elementary chirp model :  
	\begin{equation} \label{onecompmodel}
		y\left( t\right) = A^{0} e^{i\beta^{0} t^{2}}+ \epsilon\left( t\right); ~~ t=1,\dots, N.
	\end{equation}	
	Here, we need to estimate the unknown amplitude parameter and the frequency rate parameter under the following assumption on the error random variables $\epsilon\left( t\right)^\prime$s.
	
	\begin{assumption} \label{ass1}
		$\epsilon\left( t\right)^\prime$s are \textit{i.i.d.} complex-valued random variables with mean 0
		and variance $\frac{\sigma^{2}}{2}$ for both real and imaginary parts. Also, fourth order moment of $\epsilon\left( t\right)$ exists. It is assumed that real and imaginary parts of $\epsilon\left( t\right)$ are independent. 
	\end{assumption}
	
	\noindent We denote by $A_{R}$ and $A_{I}$, the real and the imaginary part of the $A$; respectively, and the real and the imaginary part of the $\epsilon\left( t\right) $ are denoted as $\epsilon_{R}\left( t\right) $ and $\epsilon_{I}\left( t\right) $; respectively. We will use the following notations: $\bm{\theta}=\left( A_{R}, A_{I}, \beta\right) $, the parameter vector,  $\bm{\theta}^{0}=\left( A_{R}^{0}, A_{I}^{0}, \beta^{0} \right) $, the true parameter vector,  $\hat{\bm{\theta}}=\left( \hat{A}_{R}, \hat{A}_{I}, \hat{\beta}\right) $, the LSE of $\bm{\theta}^{0}$ and $\tilde{\bm{\theta}}=\left( \tilde{A}_{R}, \tilde{A}_{I}, \tilde{\beta}\right) $, the ALSE of $\bm{\theta}^{0}$.

	\noindent Under the above assumption on the noise, we present two estimation techniques: the least squares estimation technique and the
	approximate least squares estimation technique, in the following subsections. We also establish the statistical properties of these estimators.
	\subsection{Least Squares Estimators} \label{onecomplse}
	
	\noindent Least squares estimation method is one of the most intutive choices to estimate the unknown parameters of the model.
	Let us denote $\bm{\Theta}_{1} = \left[-M, M \right] \times \left[-M, M \right] \times \left[0, 2\pi \right] $ as a parameter space. The assumption on the unknown parameters is mentioned below:\begin{assumption} \label{ass2}
		Let $\bm{\theta}^{0}$ be an interior point of the parameter space $\bm{\Theta}_{1}$, and $\left| A^{0} \right| >0$.
		
	\end{assumption}  The LSEs of the parameters of the model \eqref{onecompmodel} are obtained by minimizing the following residual sum of squares, say:
	\begin{equation}
		Q\left( \bm{\theta} \right) =\sum_{t=1}^{N} \left| y\left( t\right) -  A e^{i\beta t^{2}}\right|^{2},
	\end{equation}
	with respect to $ A$ and $\beta $ simultaneously, where $A= A_{R}+i A_{I}$. 
	In matrix notation, $ Q\left( \bm{\theta} \right) $ can be expressed as follows;
	\begin{equation} \label{tssonecomp}
		Q\left( \bm{\theta} \right) = \left[\bm{Y}- \bm{Z}\left(\beta\right) {A} \right]^{\mathsf{H}} \left[\bm{Y}- \bm{Z}\left( \beta\right) {A} \right],
	\end{equation}
	where $\bm{Y}_{N\times 1} =\begin{pmatrix}
		y\left( 1\right), & \dots, & y\left( N \right)
	\end{pmatrix}^{\top}$ 
	and $\bm{Z}\left( \beta\right)=	
	\begin{bmatrix}
		e^{i\beta}, & \dots, & e^{i \beta N^{2}} 
	\end{bmatrix}^{\top}.$
	
	\noindent From  \eqref{tssonecomp}, note that $A$ can be separated from $\beta$, as it is a linear parameter. Therefore, by using separable regression technique of Richards \cite{richard}, for fixed $\beta$, the LSE of $ A $ can be determined as
	\begin{equation} \label{linestonecomp}
		\hat{A}\left(\beta\right) =\left[ \bm{Z}\left( \beta\right)^{\mathsf{H}}\bm{Z}\left( \beta\right)\right]^{-1} \bm{Z}\left( \beta\right)^{\mathsf{H}}\bm{Y}.
	\end{equation}
	By replacing $A$ by $\hat{A}\left(\beta\right)$ in \eqref{tssonecomp}, we get
	\begin{equation} \label{rssonecomp}
		R\left(\beta\right) = Q\left( \hat{A}\left( \beta\right), \beta\right)  = \bm{Y}^{\mathsf{H}}\left[ \bm{I}-\bm{P_{Z}} \right] \bm{Y},
	\end{equation}
	where
	\begin{equation*}
		\bm{P_{Z}}= \bm{Z}\left( \beta\right)\left[\bm{Z}\left( \beta\right)^{\mathsf{H}}\bm{Z}\left( \beta\right)\right]^{-1} \bm{Z}\left( \beta\right)^{\mathsf{H}},
	\end{equation*}
	is the projection matrix on the column space spanned by the matrix $\bm{Z}\left( \beta\right) $. Thus, we can obtain the LSE $\hat{\beta}$ of $\beta^{0}$ by minimizing $	R\left(\beta\right)$ with respect to $\beta$. Then the LSE of $\beta$ is used to estimate the LSE of $A$, by substituting $\hat{\beta}$ in   \eqref{linestonecomp}. \\
	
	\noindent	The strong consistency and asymptotic normality of the LSEs are shown by the following results.
	
	\begin{theorem} \label{consthonecomplse}
		If assumptions \ref{ass1} and \ref{ass2} are satisfied,  then $\hat{\bm{\theta}}$ is a strongly consistent estimator of $\bm{\theta}^{0}$, i.e., \begin{center}	$\hat{\bm{\theta}} \xrightarrow{a.s.} \bm{\theta}^{0}$ as $N \rightarrow \infty$.
		\end{center}
	\end{theorem}
	\textit{Proof}	See subsection \nameref{lseonecompapp}.

	\begin{theorem} \label{asydthonecomplse}
		If assumptions \ref{ass1} and \ref{ass2} hold true, then \begin{center}	$( \hat{\bm{\theta}} - \bm{\theta}^{0}) \bm{D}^{-1} \xrightarrow{d}\mathcal{N}_3 \left( 0,  \sigma^{2}  \bm{\Sigma}^{-1} \right) $ as $N \rightarrow \infty$, 
		\end{center}
		where $\bm{D}=diag\left(\frac{1}{\sqrt{N}}, \frac{1}{\sqrt{N}}, \frac{1}{N^{2}\sqrt{N}}\right) $ and
		\begin{equation*}
			\bm{\Sigma}^{-1}=\begin{bmatrix}
				\frac{1}{2}+\frac{5A_{I}^{0^2}}{8\left|A^0 \right|^{2} } & \frac{-5A_{R}^{0} A_{I}^{0}}{8\left|A^0 \right|^{2} } & \frac{15 A_{I}^{0}}{8\left|A^0 \right|^{2}} \\
				\frac{-5A_{R}^{0} A_{I}^{0}}{8\left|A^0 \right|^{2} } & 	\frac{1}{2}+\frac{5A_{R}^{0^2}}{8\left|A^0 \right|^{2} } &  \frac{-15 A_{R}^{0}}{8\left|A^0 \right|^{2}} \\
				\frac{15 A_{I}^{0}}{8\left|A^0 \right|^{2}} & \frac{-15 A_{R}^{0}}{8\left|A^0 \right|^{2}}   & \frac{45}{8\left|A^0 \right|^{2}}\end{bmatrix}.
		\end{equation*}

	\end{theorem}
	\textit{Proof}	See subsection \nameref{lseonecompapp}.\\

	\noindent	Although LSEs have the desired theoretical asymptotic properties, obtaining the least squares estimators in practice is computationally quite challenging. For example, even for a sinusoidal model, it has been studied that the least squares surface has a number of local minima around the true parameter value (see, Rice and Rosenblatt \cite{ricerosenblatt}, for more details) and due to this reason most of the iterative methods converge to a local minimum. Therefore, any iterative procedure requires a good set of initial values (close to the true parameter value) for its convergence to the global minimum. We encounter a similar problem for the elementary chirp model as well.  Therefore, computing the LSEs for the model \eqref{onecompmodel} is also a numerically difficult problem.\\
	
		\noindent Periodogram estimators are one of the most prominent approaches for determining the initial values of the sinusoidal model's frequencies. Maximizing the following periodogram function \cite{walker} provides these estimators: 	\begin{equation} \label{pf}
		I^{0}\left( {\omega}\right)=   \frac{1}{N} \left| \sum_{t=1}^{N}  y\left( t\right) e^{-i \omega t}\right|^{2},
	\end{equation}
	over the Fourier frequencies $ \dfrac{\pi k}{N}, k=1, \dots, N-1 $. Now, we define a periodogram-type function \cite{ alsechirplike} analogous to the periodogram function which has the following mathematical form:
	\begin{equation} \label{ptf}
		I\left( \beta\right)=   \frac{1}{N} \left| \sum_{t=1}^{N}  y\left( t\right) e^{-i \beta t^{2}}\right|^{2}.
	\end{equation}
	
	\noindent	Analogous to the periodogram estimator, periodogram-type estimator is obtained by maximizing \eqref{ptf} over the grid of the type $ \dfrac{2\pi k}{N^{2}}, k=1, \dots, N^{2}-1 $, which provides the estimator of $\beta^{0}$ with the rate of convergence $ O_{P}\left(N^{-2} \right)  $. This can be used as the initial value for the frequency rate parameter.\\
	It has been established in the literature that if $I^{0}\left( {\omega}\right)$ is maximized over the continuous range $[0, \pi]$, then the obtained estimator possesses the same asymptotic properties as the corresponding LSE and hence known as ALSE \cite{walker}.
	In the next subsection, we discuss ALSEs for the elementary chirp model.
	
	\subsection{Approximate Least Squares Estimators} \label{onecompalse}
	\noindent Let us denote $\bm{\Theta}_{2} = \left( -\infty, \infty \right)  \times \left( -\infty, \infty \right) \times \left[0, 2\pi \right] $ as a parameter space. Also, the assumption on the unknown parameters is given as follows:\begin{assumption} \label{ass3}
		Let $\bm{\theta}^{0}$ be an interior point of the parameter space $\bm{\Theta}_{2}$, and $\left| A^{0} \right| >0$.
	\end{assumption}
	The ALSE of the frequency rate is obtained by maximizing the periodogram-type function  \eqref{ptf} continuously over the interval $ \left( 0, 2\pi \right)  $, that is,
	\begin{equation}\label{betalseonecomp}
		\tilde{\beta} =\operatorname{arg}\underset{\beta}{\operatorname{max}}~I\left(\beta \right).
	\end{equation} 
	
	\noindent The periodogram-type function \eqref{ptf} can also be expressed as follows:
	\begin{equation}\label{ptfexp}
		\begin{split}
			I\left(\beta \right) &=\frac{1}{N} \left\lbrace\sum_{t=1}^{N} \left( y_{R}\left( t\right) \cos\left( \beta t^{2}\right) +  y_{I}\left( t\right) \sin\left( \beta t^{2}\right) \right)  \right\rbrace^{2}\\ &+\frac{1}{N}\left\lbrace\sum_{t=1}^{N} \left( y_{I}\left( t\right) \cos\left( \beta t^{2}\right) -  y_{R}\left( t\right) \sin\left( \beta t^{2}\right) \right) \right\rbrace^{2} .
		\end{split}
	\end{equation}
	Here, $y_{R}\left( t\right) $ and $y_{I}\left( t\right)$ are the real and imaginary parts of the model \eqref{onecompmodel}, respectively, and are expressed as follows:
	\begin{equation} \label{yronecompmodelexp}
		y_{R}\left( t\right)= A_{R}^{0} \cos\left(\beta^{0} t^{2} \right) -A_{I}^{0} \sin\left( \beta^{0} t^{2} \right) +\epsilon_{R}\left( t\right),  
	\end{equation}
	\begin{equation} \label{yionecompmodelexp}
		y_{I}\left( t\right)= A_{R}^{0} \sin\left(\beta^{0} t^{2} \right) +A_{I}^{0} \cos\left( \beta^{0} t^{2} \right) +\epsilon_{I}\left( t\right). 
	\end{equation}
	Once we obtain $\tilde{\beta}$, then the  ALSEs of the linear parameters $\tilde{A}_{R}$ and $\tilde{A}_{I}$ can be determined using the technique of simple linear regression and expressed as follows:
	\begin{equation} \label{aralseonecomp}
		\tilde{A}_{R} = \frac{1}{N} \sum_{t=1}^{N} \left(y_{R} \cos\left( \tilde{\beta}t^{2}\right)+ y_{I} \sin\left( \tilde{\beta}t^{2}\right) \right),
	\end{equation}
	\begin{equation} \label{aialseonecomp}
		\tilde{A}_{I} = \frac{1}{N} \sum_{t=1}^{N} \left(y_{I} \cos\left( \tilde{\beta}t^{2}\right)- y_{R} \sin\left( \tilde{\beta}t^{2}\right) \right).
	\end{equation}
	
	\noindent In the following theorems, we present the strong consistency and the asymptotic distribution results of the ALSEs.
	\begin{theorem}  \label{consthonecompalse}
		If assumptions \ref{ass1} and \ref{ass3} hold true, then $\tilde{\bm{\theta}}$ is a strongly consistent estimator of $\bm{\theta}^{0}$, i.e., \begin{center}	$\tilde{\bm{\theta}} \xrightarrow{a.s.} \bm{\theta}^{0}$ as $N \rightarrow \infty$.
		\end{center}
	\end{theorem}
	\textit{Proof}	See subsection \nameref{alseonecompapp}.
	
	\begin{theorem} \label{asydthonecompalse}
		If assumptions \ref{ass1} and \ref{ass3} hold true, then the asymptotic distribution of $( \tilde{\bm{\theta}} - \bm{\theta}^{0}) \bm{D}^{-1}$ is identical to the asymptotic distribution of $( \hat{\bm{\theta}} - \bm{\theta}^{0}) \bm{D}^{-1}$ as $N \rightarrow \infty$,
		where $\bm{D}=diag\left(\frac{1}{\sqrt{N}}, \frac{1}{\sqrt{N}}, \frac{1}{N^{2}\sqrt{N}}\right) $.
	\end{theorem}
	\textit{Proof}	See subsection \nameref{alseonecompapp}.\\
	
	\noindent The obtained ALSEs achieve the optimal rates of convergence and are also asymptotically identical to their corresponding LSEs. Therefore, the obtained estimators are called the  ALSEs. Also, note that to prove the asymptotic theoretical results of the ALSEs, we need somewhat weaker assumptions on the parameter space than those needed for the LSEs.
	
	\section{Multiple-Component Elementary Chirp model} \label{multicompsec}
	\noindent In this section, we consider a multiple-component elementary
	chirp model \eqref{multicompmodel}. Along with the assumptions on error random variables, certain assumptions on the parameters are needed to prove the asymptotic theoretical properties of the LSEs and sequential LSEs, which are stated below. In the following subsections, we provide the theoretical results under the stated assumptions. 
	
	\noindent Let us denote $ {\scriptsize{\bm{v}}} $ as the parameter vector for the model \eqref{multicompmodel}, ${\scriptsize{\bm{v}}}=\left(A_{R1}, A_{I1}, \beta_{1}, \dots, A_{Rp}, A_{Ip}, \beta_{p} \right) $. Also, denote ${\scriptsize{\bm{v}}}^{0}$ as the true parameter vector, $\hat{{\scriptsize{\bm{v}}}}$ as the LSE of ${\scriptsize{\bm{v}}}^{0}$, $\breve{{\scriptsize{\bm{v}}}}$ as the sequential LSE of ${\scriptsize{\bm{v}}}^{0}$ and $\tilde{{\scriptsize{\bm{v}}}}$ as the sequential ALSE of ${\scriptsize{\bm{v}}}^{0}$.
	\begin{assumption} \label{ass4}
		Let ${\scriptsize{\bm{v}}}^{0}$ be an interior point of the parameter space $ \bm{\mathcal{V}}_{1}=\bm{\Theta}_{1}^{\left(p \right) }$; 	$\bm{\Theta}_{1} = \left[-M, M \right] \times \left[-M, M \right] \times \left[0, 2\pi \right]$ and the frequency rates $\beta_{k}^{0}$s are distinct for $k=1, \cdots, p$.
	\end{assumption}
	\begin{assumption}\label{ass5}
		The amplitude parameters; $A_{k}^{0}$s satisfy the following relationship:
		\begin{equation*}
			2M^{2}>	\left| A_{1}^{0} \right| ^{2} >\left| A_{2}^{0} \right| ^{2} >\dots>\left| A_{p}^{0} \right| ^{2} >0.
		\end{equation*}
	\end{assumption}
	\subsection{Least Squares Estimators}
	\noindent The LSEs of the unknown parameters of the multiple-component elementary chirp model \eqref{multicompmodel} can be obtained by minimizing the following residual sum of squares:
	\begin{equation}
		Q\left( {\scriptsize{\bm{v}}}\right) =\sum_{t=1}^{N} \left| y\left( t\right) -  \sum_{k=1}^{p} A_{k} e^{i\beta_{k} t^{2}}\right|^{2},
	\end{equation}
	with respect to $ A_{1}, \beta_{1}, \dots, A_{p}$ and $\beta_{p} $ simultaneously. Now the LSE, $\hat{{\scriptsize{\bm{v}}}}$ of ${\scriptsize{\bm{v}}}^{0}$, can be obtained in the similar manner as done in the
	one-component elementary chirp model.
	Now, we present the strong consistency and the asymptotic distribution results of the LSEs.
	\begin{theorem} \label{consthmulticomplse}
		If assumptions \ref{ass1}, \ref{ass4} and \ref{ass5} hold true, then $\hat{{\scriptsize{\bm{v}}}}$ is a strongly consistent estimator of ${\scriptsize{\bm{v}}}^{0}$, i.e., \begin{center}	$\hat{{\scriptsize{\bm{v}}}} \xrightarrow{a.s.} {\scriptsize{\bm{v}}}^{0}$ as $N \rightarrow \infty$.
		\end{center}
	\end{theorem}
	\textit{Proof}	This result can be proved in the similar manner to Theorem \ref{consthonecomplse}.

	\begin{theorem} \label{asydthmulticomplse}
		If assumptions \ref{ass1}, \ref{ass4} and \ref{ass5} are satisfied, then \begin{center}	$( \hat{{\scriptsize{\bm{v}}}} - {\scriptsize{\bm{v}}}^{0}) \bm{\mathcal{D}}^{-1} \xrightarrow{d}\mathcal{N}_{3p} \left( 0,  \sigma^{2}  \mathcal{E}^{-1} \right) $ as $N \rightarrow \infty$.
		\end{center}
		Here, $\bm{\mathcal{D}}=diag(\underset{p~times}{\underbrace{\bm{D}, \dots, \bm{D}}})$, where $ \bm{D}=diag\left(\frac{1}{\sqrt{N}}, \frac{1}{\sqrt{N}}, \frac{1}{N^{2}\sqrt{N}}\right) $ and
		\begin{center}
			$
			\mathcal{E}^{-1}=\begin{bmatrix}
				\Sigma_{1}^{-1} &0 & \dots & 0\\
				0&\Sigma_{2}^{-1}&\dots &0\\
				\vdots & \vdots &\ddots & \vdots \\
				0&0& 0 & \Sigma_{p}^{-1}
			\end{bmatrix}$,
			
		\end{center}
		with 
		\begin{equation} \label{sigmainv}
			\bm{\Sigma}_{k}^{-1}=\begin{bmatrix}
				\frac{1}{2}+\frac{5A_{Ik}^{0^2}}{8\left|A_{k}^0 \right|^{2} } & \frac{-5A_{Rk}^{0} A_{Ik}^{0}}{8\left|A_{k}^0 \right|^{2} } & \frac{15 A_{Ik}^{0}}{8\left|A_{k}^0 \right|^{2}} \\
				\frac{-5A_{Rk}^{0} A_{Ik}^{0}}{8\left|A_{k}^0 \right|^{2} } & 	\frac{1}{2}+\frac{5A_{Rk}^{0^2}}{8\left|A_{k}^0 \right|^{2} } &  \frac{-15 A_{Rk}^{0}}{8\left|A_{k}^0 \right|^{2}} \\
				\frac{15 A_{Ik}^{0}}{8\left|A_{k}^0 \right|^{2}} & \frac{-15 A_{Rk}^{0}}{8\left|A_{k}^0 \right|^{2}}   & \frac{45}{8\left|A_{k}^0 \right|^{2}} 
			\end{bmatrix}.
		\end{equation}
	\end{theorem}
	\textit{Proof}	See subsection \nameref{lsemulticompapp}.

	\subsection{Choice of initial values}
\begin{figure}[!t]
		\includegraphics[width=\linewidth, height=8cm]{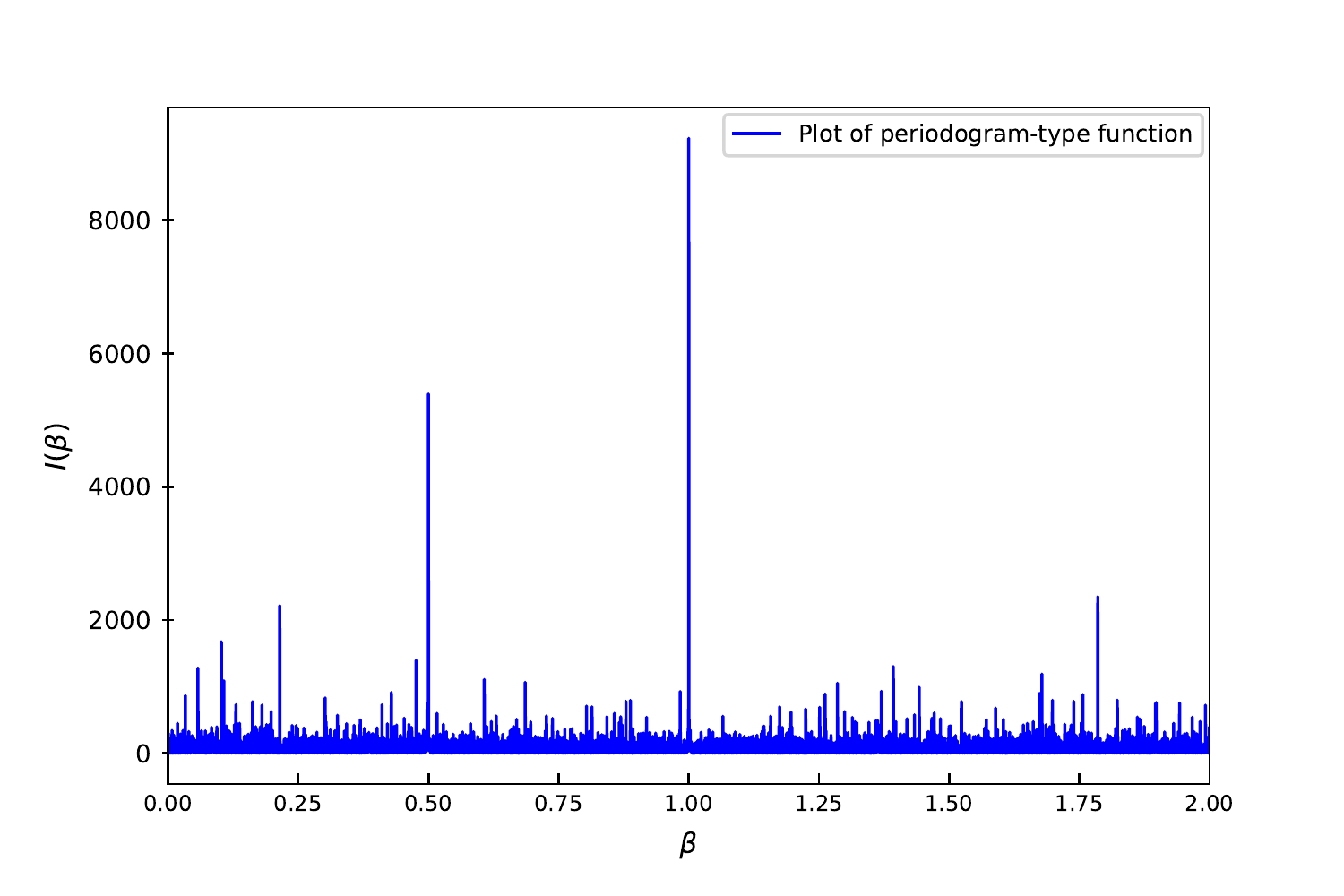}
		\caption{ Plot of the periodogram-type function \eqref{ptf} of the data obtained by \eqref{2compmodel}.}
		\label{ptfplot}
	\end{figure}

	\noindent In all these methods discussed in section (\ref{onecompsec}), we use the periodogram-type estimator as an initial value. Also, to find the LSEs for the multiple-component elementary chirp model, periodogram-type estimators are used as the initial values, which are the values of frequency rates where we get the peaks of the periodogram-type function \eqref{ptf}. Let us see this through an illustration for the following two-component elementary chirp model
	\begin{equation} \label{2compmodel}
		y\left( t\right) = y_{1}\left( t\right) =7 e^{1i t^{2}}+  5 e^{0.5i t^{2}}+\epsilon\left( t\right); ~~ t=1,\dots, 200.
	\end{equation}	
	Here, $\epsilon\left(t \right)^\prime$s are i.i.d. complex-valued normal random variables and satisfy assumption \ref{ass1}. The value of $\sigma^{2}$ is $ 1 $. We generate a signal using \eqref{2compmodel}, where two frequency rates are 1 and 0.5, which are far apart. From Fig. \ref{ptfplot}, it is clear that we obtain the two highest peaks near the true frequency rates, which are clearly resolvable.
	But, in some situations, even for $p = 2$, it might be difficult to get peaks in the periodogram-type function around the true frequency rates. It may lead to initial values which are not close to the true parameter values and hence results in incorrect estimates of the parameters. To demonstrate such situations, we take two different scenarios:\\
	Case 1: When two frequency rates are close to each other.
	\begin{equation} \label{2qcbmodel}
		y\left( t\right) = y_{1}\left( t\right) =2 e^{1.45i t^{2}}+  1 e^{1.5i t^{2}}+\epsilon\left( t\right); ~ t=1,\dots, 100.
	\end{equation}	
	\begin{equation} \label{2qcbmodelupd}
		y_{2}\left( t\right) =	y_{1}\left( t\right) -\breve{A}_{1} e^{i\breve{\beta}_{1}t^{2}}; ~ t=1,\dots, 100.
	\end{equation}	
	Case 2: When two frequency rates are far apart but error variance is high.
	\begin{equation} \label{hevmodel}
		y\left( t\right) =y_{1}\left( t\right) = 4 e^{2i t^{2}}+ 2 e^{1.5i t^{2}}+\epsilon\left( t\right); ~ t=1,\dots, 100.
	\end{equation}	
	\begin{equation} \label{hevmodelupd}
		y_{2}\left( t\right) =	y_{1}\left( t\right) -\breve{A}_{1} e^{i\breve{\beta}_{1}t^{2}}; ~ t=1,\dots, 100.
	\end{equation}	
	
	\begin{figure}[!t]
		\includegraphics[width=\linewidth, height=8cm]{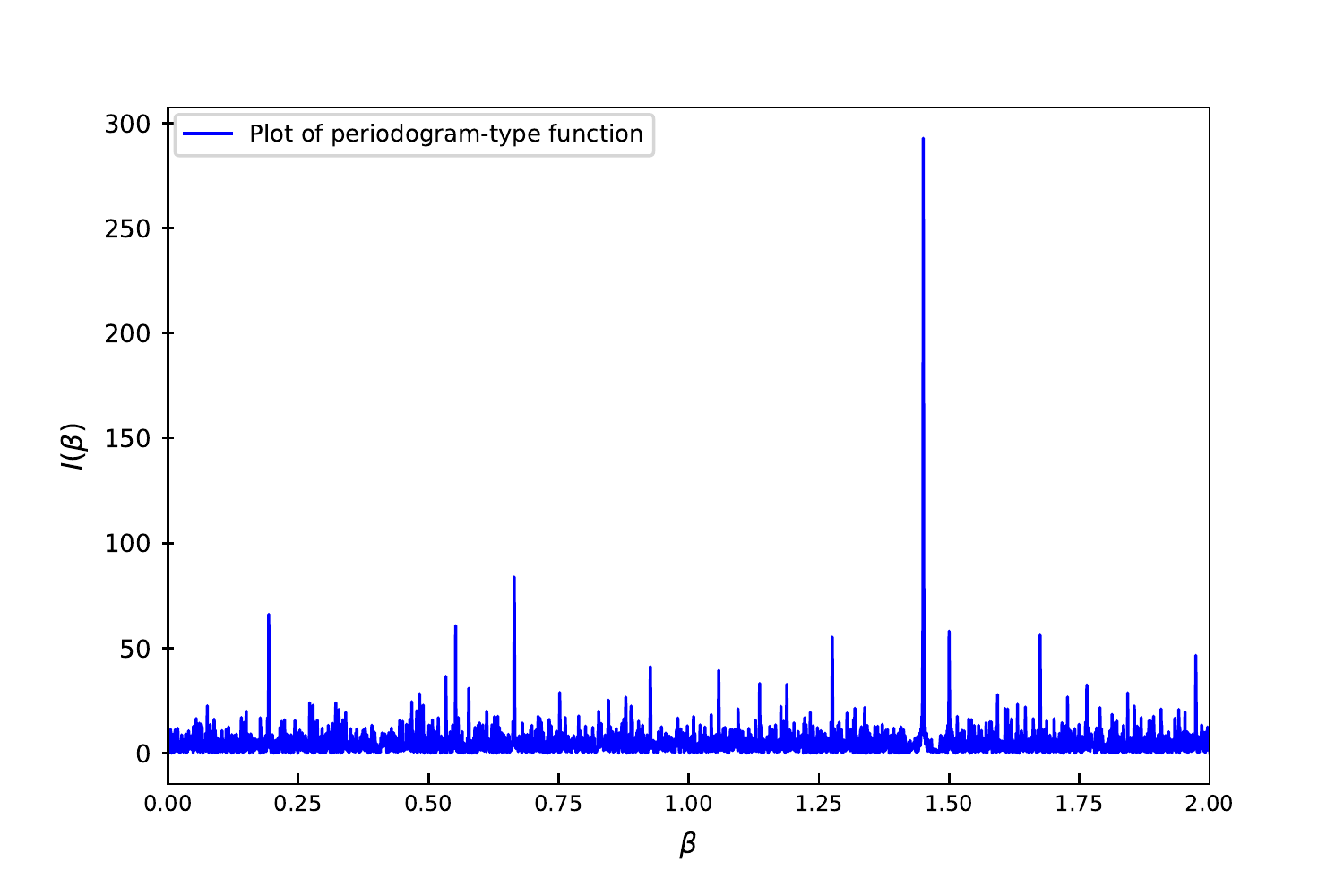}
		\caption{Plot of the periodogram-type function \eqref{ptf} of the data obtained by \eqref{2qcbmodel}.}
		\label{2qcbplot}
	\end{figure}	
	\begin{figure}[!t]
		\includegraphics[width=\linewidth, height=8cm]{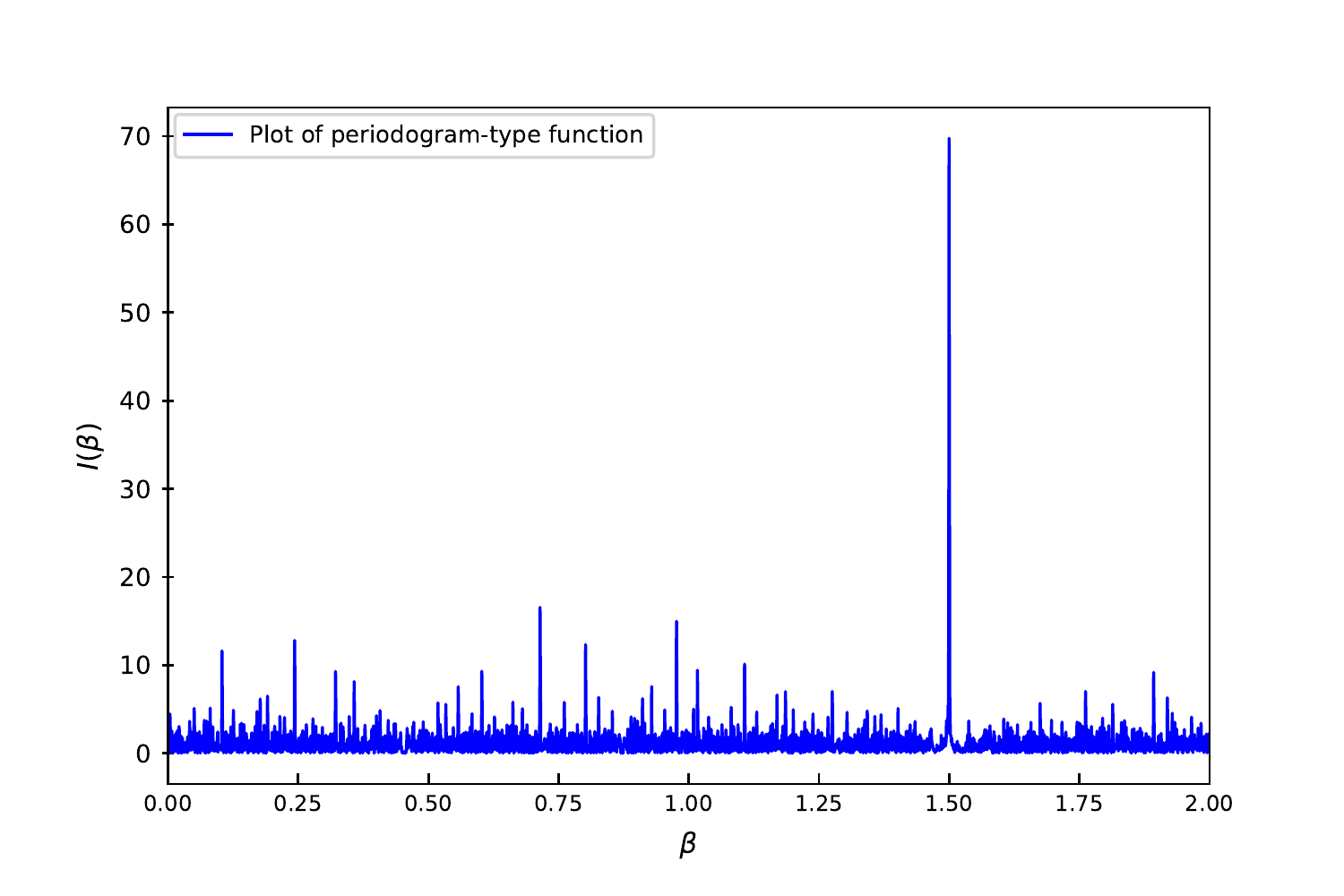}
		\caption{Plot of the periodogram-type function \eqref{ptf} of the updated data obtained by \eqref{2qcbmodelupd}.}
		\label{2qcbplotupd}
	\end{figure}	
	\begin{figure}[!t]
		\includegraphics[width=\linewidth, height=8cm]{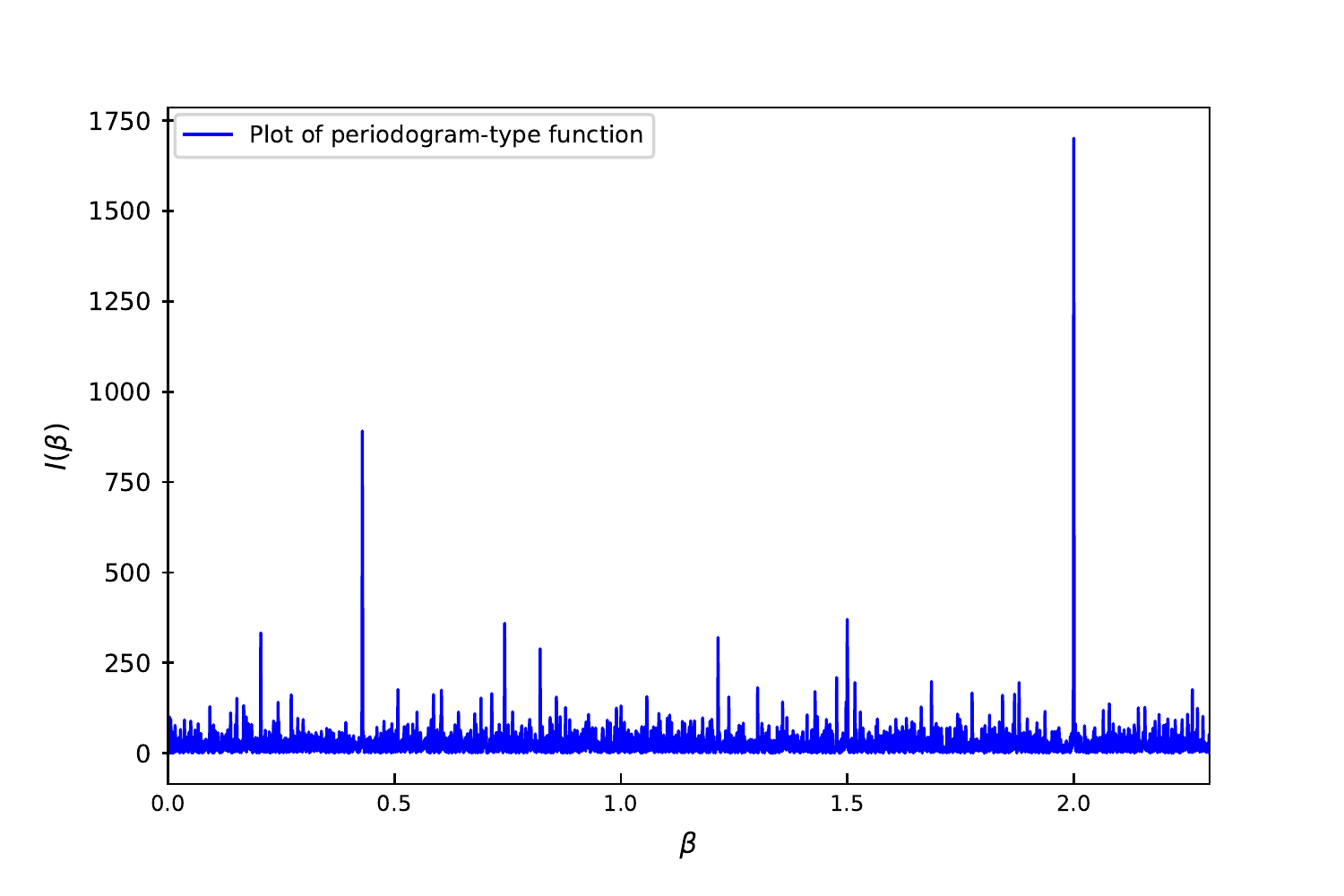}
		\caption{Plot of the periodogram-type function \eqref{ptf} of the data obtained by \eqref{hevmodel}.}
		\label{hevplot}
	\end{figure}	
	\begin{figure}[!t]
		\includegraphics[width=\linewidth, height=8cm]{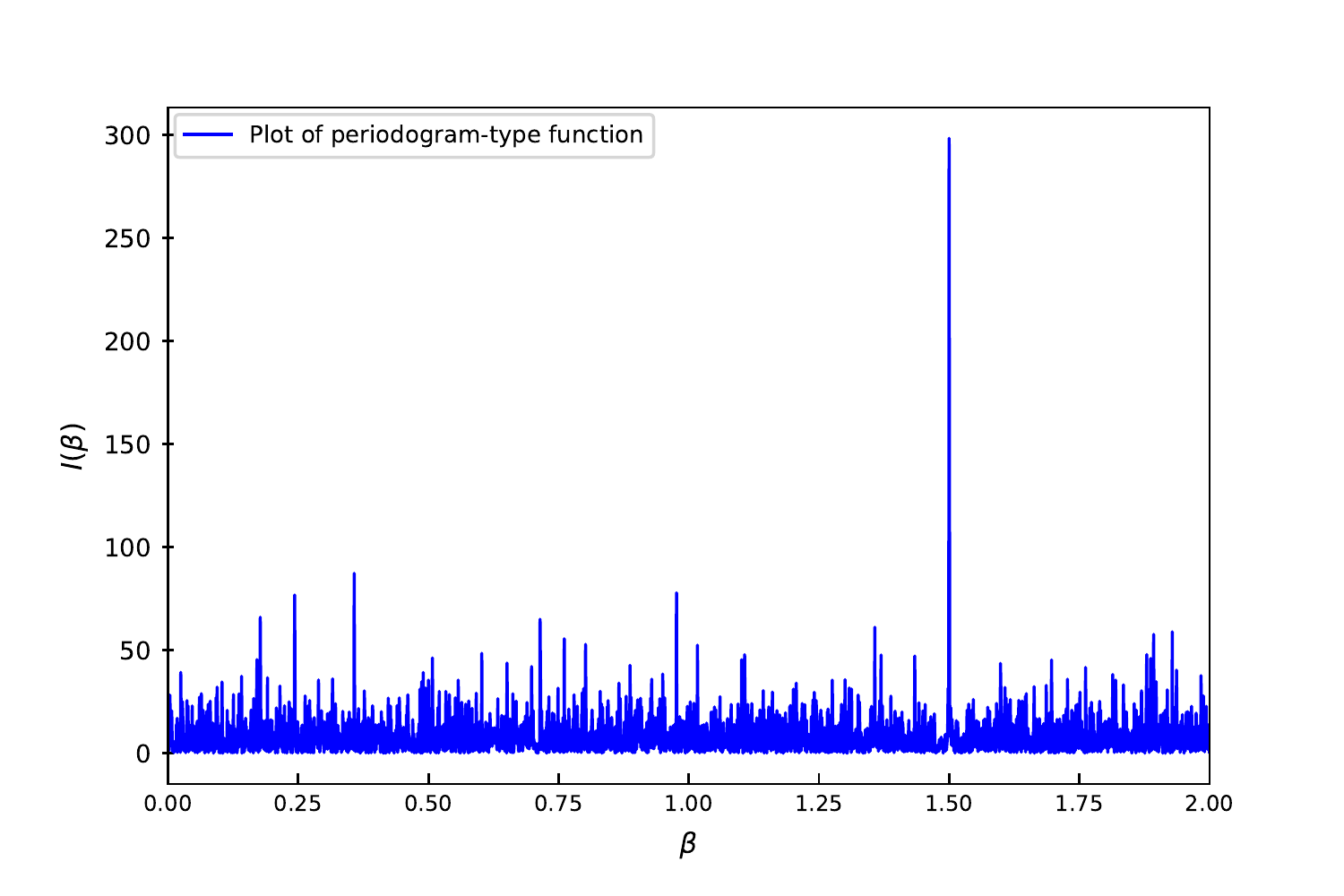}
		\caption{Plot of the periodogram-type function \eqref{ptf} of the updated data obtained by \eqref{hevmodelupd}.}
		\label{hevplotupd}
	\end{figure}
	
	\noindent The values of $\sigma^{2}$ are taken as $ 0.1 $ and $ 3 $, in case 1 and in case 2, respectively. $\breve{A}_{1}$ and $\breve{\beta}_{1}$ are the estimators of the $A_{1}^{0}$ and $\beta_{1}^{0}$, respectively. We generate a signal using \eqref{2qcbmodel}, where two frequency rates are 1.45 and 1.5, which are close to each other. Another signal is generated using \eqref{hevmodel}, where two frequency rates are well separated, but the error variance is relatively high. When we plot the periodogram-type function for these signals, we are unable to get the two highest peaks near the true frequency rates, see Fig. \ref{2qcbplot} and Fig. \ref{hevplot}. To solve this problem, we use one of the sequential procedures presented in the next two subsections. It is observed from Fig. \ref{2qcbplot} and Fig. \ref{hevplot} that we are able to detect the correct peak of one of the components, having the highest amplitude for both models. So using these initial values, we estimate the first component of both the models using the discussed methodologies, namely, LSEs or ALSEs. Further, we update our data by eliminating the  effect of the first estimated component using \eqref{2qcbmodelupd} and \eqref{hevmodelupd}. Then, we plot the periodogram-type function for the updated data. We see in Fig. \ref{2qcbplotupd} and Fig. \ref{hevplotupd} that peaks for the second component of both the models are obtained near the true frequency rates. Then we obtain the estimates of the second components of both models by using the updated data. \\
	In cases where periodogram-type functions are able to correctly resolve the frequency rates, computing the LSEs becomes numerically challenging, when $p$, the number of components, is large. To resolve this issue, in the next subsections, we propose sequential procedures to estimate the parameters of the model \eqref{multicompmodel}.

	\subsection{Sequential Least Squares Estimators} \label{seqlsesec}
	\noindent In this subsection, we propose a sequential method to estimate the model \eqref{multicompmodel} parameters. The $p$-dimensional optimization problem can be reduced to $p$, 1-D optimization problems using the sequential technique. This is possible because of the orthogonal structure of the chirp components which can be proved using lemma 2 of \cite{lsechirplike}. Thus, by using the orthogonality of the chirp components of the model \eqref{multicompmodel}, we lower the computational complexity of calculating the LSEs without losing the efficiency of the estimators. The procedure for the sequential method is presented in the following steps:\\
	
	\noindent \textbf{Step 1:}  Obtain the estimate of the parameters of the first chirp component, i.e., $\breve{\bm{\theta}}_{1}=\left( \breve{A_{R1} }, \breve{A_{I1} }, \breve{\beta}_{1} \right) $  of the multiple-component elementary chirp model \eqref{multicompmodel}, using the method described in subsection \ref{onecomplse} for the one-component elementary chirp model \eqref{onecompmodel}.\\
	
	\noindent	\textbf{Step 2:} Take out the effect of the estimated chirp component and obtain the adjusted data vector: \begin{equation*}
		y_{2}\left(t \right) =y_{1}\left( t\right) - \breve{A}_{1} e^{i\breve{\beta}_{1}t^{2}}; ~~~ y_{1}\left( t\right) =y\left( t\right).
	\end{equation*}

	\noindent	\textbf{Step 3:} To estimate the parameters of the second chirp component i.e. $\breve{\bm{\theta}}_{2 } =\left( \breve{A}_{R2}, \breve{A}_{I2}, \breve{\beta}_{2} \right) $, minimize the following residual sum of squares: 
	\begin{equation*}
		Q_{2}\left( \bm{\theta} \right) =\sum_{t=1}^{N} \left| y_{2}\left( t\right) -  A e^{i\beta t^{2}}\right|^{2}.
	\end{equation*}
	Repeat this procedure until all the $p$-chirp components are estimated.\\
	
	\noindent We now provide the results for the strong consistency of the proposed sequential LSEs, when $ p $ is unknown. Therefore, we take the following situations: (1) when the fitted model's number of components is less than or equal to the true number of components, and (2) when the fitted model's number of components is more than the true number of components.
	
	\begin{theorem}\label{consthseqlse1}
		If assumptions \ref{ass1},  \ref{ass4} and \ref{ass5} hold true, then $\breve{\bm{\theta}}_{1}$ is a strongly consistent estimator of $\bm{\theta}_{1}^{0}$, i.e., \begin{center}	$\breve{\bm{\theta}}_{1} \xrightarrow{a.s.} \bm{\theta}_{1}^{0}$ as $N \rightarrow \infty$.
		\end{center}
	\end{theorem}
	\textit{Proof}	See subsection \nameref{seqlseapp}.
	
	\begin{theorem}\label{consthseqlse2}
		If assumptions \ref{ass1},  \ref{ass4} and \ref{ass5} hold true, then $\breve{\bm{\theta}}_{2}$ is a strongly consistent estimator of $\bm{\theta}_{2}^{0}$, i.e., \begin{center}	$\breve{\bm{\theta}}_{2} \xrightarrow{a.s.} \bm{\theta}_{2}^{0}$ as $N \rightarrow \infty$.
		\end{center}
	\end{theorem}
	\textit{Proof}	See subsection \nameref{seqlseapp}.
	
	This theorem can also be extended for $k\leq p$.
	\begin{theorem}
		If assumptions \ref{ass1},  \ref{ass4} and \ref{ass5} hold true, then $\breve{\bm{\theta}}_{k}$ is a strongly consistent estimator of $\bm{\theta}_{k}^{0}$, i.e., \begin{center}	$\breve{\bm{\theta}}_{k} \xrightarrow{a.s.} \bm{\theta}_{k}^{0}$ as $N \rightarrow \infty$, $ \forall k\leq p$.
		\end{center}
	\end{theorem}
	\textit{Proof}	This theorem can be proved using the similar argument as in theorem \ref{consthseqlse2}.

	\begin{theorem}\label{consthseqlse3}
		If assumptions \ref{ass1},  \ref{ass4} and \ref{ass5} hold true, then  \begin{center}	$\breve{A}_{R\left(p+k \right) } \xrightarrow{a.s.} 0$, $\breve{A}_{I\left(p+k \right) } \xrightarrow{a.s.} 0$ as $N \rightarrow \infty$, $\forall k = 1, 2, \cdots $.
		\end{center}
	\end{theorem}
	\textit{Proof}	See subsection \nameref{seqlseapp}.

	\noindent The asymptotic distribution of the proposed sequential estimators is given by the following
	theorem:
	\begin{theorem} \label{asydthseqlse1}
		If assumptions \ref{ass1},  \ref{ass4} and \ref{ass5} hold true, then \begin{center}	$( \breve{\bm{\theta}}_{k} - \bm{\theta}_{k}^{0}) \bm{D}^{-1} \xrightarrow{d}\mathcal{N}_3 \left( 0,  \sigma^{2}  \bm{\Sigma}_{k}^{-1} \right) $ as $N \rightarrow \infty$, $ \forall k\leq p$, 
		\end{center}
		where $\bm{D}=diag\left(\frac{1}{\sqrt{N}}, \frac{1}{\sqrt{N}}, \frac{1}{N^{2}\sqrt{N}}\right) $ and $\bm{\Sigma}_{k}^{-1}$ is same as defined in \eqref{sigmainv}.
	\end{theorem}
	\textit{Proof}	See subsection \nameref{seqlseapp}.
	
	\subsection{Sequential Approximate Least Squares Estimators }
	\noindent The following assumptions are required to prove the asymptotic properties of sequential ALSEs. 
	\begin{assumption} \label{ass6}
		Let ${\scriptsize{\bm{v}}}^{0}$ be an interior point of the parameter space $ \bm{\mathcal{V}}_{2}=\bm{\Theta}_{2}^{ \left( p\right)  }$; 	$\bm{\Theta}_{2} = \left( -\infty, \infty \right)  \times \left( -\infty, \infty \right) \times \left[0, 2\pi \right] $ and the frequency rates $\beta_{k}^{0}$s are distinct for $k=1, \cdots, p$.
	\end{assumption}
	\begin{assumption}\label{ass7}
		The amplitude parameters; $A_{k}^{0}$s satisfy the following relationship:
		\begin{equation*}
			\infty>	\left| A_{1}^{0} \right|  >\left| A_{2}^{0} \right|  >\dots>\left| A_{p}^{0} \right|  >0.
		\end{equation*}
	\end{assumption}
	
	\noindent The algorithm for the sequential ALSEs is described in the following steps:\\
	
	\noindent \textbf{Step 1:} Obtain the estimate of the parameters of the first chirp component, i.e., $\tilde{\bm{\theta}}_{1}=\left( \tilde{A }_{R1}, \tilde{A }_{I1}, \tilde{\beta}_{1} \right) $  of the multiple-component elementary chirp model \eqref{multicompmodel}, using the method described in subsection  \ref{onecompalse} for the one-component elementary chirp model \eqref{onecompmodel}.\\
	
	\noindent	\textbf{Step 2:} Take out the effect of the estimated chirp component and obtain the new data vector: \begin{equation*}
		y_{2}\left(t \right) =y_{1}\left( t\right) - \tilde{A}_{1} e^{i\tilde{\beta}_{1}t^{2}}; ~~~ y_{1}\left( t\right) =y\left( t\right).
	\end{equation*}

	\noindent	\textbf{Step 3:} Maximize the following periodogram-type function to compute $\tilde{\beta}_{2}$: 
	\begin{equation*}
		I_{2}\left( \beta\right)=   \frac{1}{N} \left| \sum_{t=1}^{N}  y_{2}\left( t\right) e^{-i \beta t^{2}}\right|^{2}.
	\end{equation*}
	Substitute $\tilde{\beta}_{2}$, $y_{R2}\left(t \right) $ and $y_{I2}\left(t \right) $ in \eqref{aralseonecomp} and \eqref{aialseonecomp}, to obtain $ \tilde{A}_{R2}$ and $\tilde{A}_{I2} $. Thus, we obtain the estimate of the second chirp component.
	Repeat this procedure until all the $p$-chirp components are estimated.\\

	\noindent We now provide the results for the strong consistency of the proposed sequential ALSEs, when $ p $ is unknown. Therefore, we take the following situations: (1) when the fitted model's number of components is less than or equal to the true number of components, and (2) when the fitted model's number of components is more than the true number of components.
	\begin{theorem}\label{consthseqalse1}
		If assumptions \ref{ass1},  \ref{ass6} and \ref{ass7} hold true, then $\tilde{\bm{\theta}}_{1}$ is a strongly consistent estimator of $\bm{\theta}_{1}^{0}$, i.e., \begin{center}	$\tilde{\bm{\theta}}_{1} \xrightarrow{a.s.} \bm{\theta}_{1}^{0}$ as $N \rightarrow \infty$.
		\end{center}
	\end{theorem}
	\textit{Proof}	See subsection \nameref{seqalseapp}.
	
	\begin{theorem}\label{consthseqalse2}
		If assumptions \ref{ass1},  \ref{ass6} and \ref{ass7} hold true, then $\tilde{\bm{\theta}}_{2}$ is a strongly consistent estimator of $\bm{\theta}_{2}^{0}$, i.e., \begin{center}	$\tilde{\bm{\theta}}_{2} \xrightarrow{a.s.} \bm{\theta}_{2}^{0}$ as $N \rightarrow \infty$.
		\end{center}
	\end{theorem}
	\textit{Proof}	See subsection \nameref{seqalseapp}.
	
	This theorem can also be extended for $k\leq p$.
	\begin{theorem}\label{consthseqalse3}
		If assumptions \ref{ass1},  \ref{ass6} and \ref{ass7} hold true, then $\tilde{\bm{\theta}}_{k}$ is a strongly consistent estimator of $\bm{\theta}_{k}^{0}$, i.e., \begin{center}	$\tilde{\bm{\theta}}_{k} \xrightarrow{a.s.} \bm{\theta}_{k}^{0}$ as $N \rightarrow \infty$, $ \forall k\leq p$.
		\end{center}
	\end{theorem}
	\textit{Proof}	This theorem can be proved using the similar argument as in theorem \ref{consthseqalse2}.
	
	\begin{theorem}\label{consthseqalse4}
		If assumptions \ref{ass1},  \ref{ass6} and \ref{ass7} hold true, then \begin{center}	$\tilde{A}_{R\left(p+k \right) } \xrightarrow{a.s.} 0$, $\tilde{A}_{I\left(p+k \right) } \xrightarrow{a.s.} 0$ as $N \rightarrow \infty$, $\forall k = 1, 2, \cdots $.
		\end{center}
	\end{theorem}
	\textit{Proof}	See subsection \nameref{seqalseapp}.
	
	Next, using the following theorem, we obtain the asymptotic distribution of the proposed sequential ALSEs:
	\begin{theorem} \label{asydthseqalse1}
		If assumptions \ref{ass1},  \ref{ass6} and \ref{ass7} hold true, then the asymptotic distribution of $( \tilde{\bm{\theta}}_{k} - \bm{\theta}_{k}^{0}) \bm{D}^{-1}$ is identical to the asymptotic distribution of the $( \breve{\bm{\theta}}_{k} - \bm{\theta}_{k}^{0}) \bm{D}^{-1}$, $\forall k=1,\dots, p$,
		where $\bm{D}=diag\left(\frac{1}{\sqrt{N}}, \frac{1}{\sqrt{N}}, \frac{1}{N^{2}\sqrt{N}}\right) $.
	\end{theorem}
	\textit{Proof}	See subsection \nameref{seqalseapp}.\\
	
	\noindent From the above discussion, it follows that the sequential LSEs and the sequential ALSEs are  strongly consistent and have the asymptotic distribution  same as that of the LSEs. Also, they can be computed with lower computational complexity. For the implementation purpose, we use the sequential LSEs and the sequential ALSEs, to analyze the considered data.\\
	
	\noindent \textbf{Remark:} All the proposed estimation methods can be easily extended when the errors come from a stationary linear process, i.e., \begin{equation}
		\epsilon(t) = \sum_{l=-\infty}^{\infty} c\left( l\right) e\left( t-l\right) .
	\end{equation} 
	Here, $\left\lbrace e\left(t \right)\right\rbrace  $ is a sequence of \textit{i.i.d.} random variables with mean $0$, variance $\sigma^{2}$ and finite fourth moment and $c\left( l\right)^\prime$s are real numbers such that $\sum_{l=-\infty}^{\infty} \left| c\left( l\right)\right| < \infty$.
	
	\section{Numerical Experiments} \label{simdatasec}
	\noindent In this section, we present some results obtained from the simulation studies for a one-component and for a multiple-component complex-valued elementary chirp model. Here, we perform numerical experiments for different sample sizes $N$, for varying signal-to-noise ratios (SNRs) and for varying error variances $ \sigma^{2}$ and see how the proposed methodologies work and to compare their performance with some other existing estimation procedures. Here, we are mainly interested in estimating the non-linear parameters since it is straightforward to get the estimates of the linear parameters using simple linear regression once we have the estimates of the non-linear parameters. Therefore, the estimates of the non-linear parameters are provided only.
	
	\subsection{Simulation results for a one-component elementary chirp model} \label{sim1}
	\noindent We consider a one-component elementary chirp model \eqref{onecompmodel} for this simulation study with the true parameter values given as follows:
	$A^{0}=5$ and $\beta^{0}=0.5$.
	
	\noindent Here, $\epsilon\left(t \right)^\prime$s are \textit{i.i.d.} complex-valued normal random variables and satisfy assumption \ref{ass1}. We have taken $N=101, 201, 301, 401,$ and $501$ and $ \sigma^{2}= 1, 2$ and $3$. We estimate the non-linear parameter using LSEs, ALSEs, dechirping method \cite{dechirp} and CPF method \cite{cpf2}. The dechirping method is implemented in the following manner. \\
	Consider the complex conjugate of model \eqref{onecompmodel} with lag one which can be written in the following expression:
	\begin{equation} \label{lagonecompmodel}
		\overline{y\left( t+1\right)} = \overline{A^{0}} e^{-i\beta^{0} \left( t+1\right) ^{2}}+ \overline{\epsilon\left( t+1\right)}; ~~ t=1,\dots, N-1.
	\end{equation}	
	\begin{equation} \label{lagonedechirpmodel}
		\begin{split}
			&y\left( t\right)\overline{y\left( t+1\right)}  = |A^{0}|^{2} e^{-i\beta^{0} (2t+1)}+ A^{0} e^{i\beta^{0} t^{2}}\overline{\epsilon\left( t+1\right)}\\ &+\overline{A^{0}} e^{-i\beta^{0} \left( t+1\right) ^{2}}\epsilon\left( t\right) +\epsilon\left( t\right)\overline{\epsilon\left( t+1\right)}; ~~ t=1,\dots, N-1.
		\end{split}
	\end{equation}	
	Now, \eqref{lagonedechirpmodel} can be rewritten in the following expression:
	\begin{equation} \label{lagonedechitppmodel2}
		z\left( t\right) = B^{0} e^{-i2\beta^{0}t}+\epsilon_{1}\left( t\right) ; ~~ t=1,\dots, N-1;
	\end{equation}	
	where, $B^{0}=|A^{0}|^{2} e^{-i\beta^{0}}$ and $\epsilon_{1}\left( t\right) = A^{0} e^{i\beta^{0} t^{2}}\overline{\epsilon\left( t+1\right)}+\overline{A^{0}} e^{-i\beta^{0} \left( t+1\right) ^{2}}\epsilon\left( t\right) +\epsilon\left( t\right)\overline{\epsilon\left( t+1\right)}$.
	From equation \eqref{lagonedechitppmodel2}, it is clear that this is a sinusoidal model with the frequency parameter $-2\beta^{0}$ and amplitude parameter $B^{0}$. Then the LSEs of $B^{0}$ and $\beta^{0}$ are determined by minimizing the following error sum of squares:
	\begin{equation} \label{objfundechirp}
		Q\left( {B,\beta} \right) =\sum_{t=1}^{N-1} \left| z\left( t\right) -  B e^{-i2\beta t}\right|^{2},
	\end{equation}
	with respect to $B$ and $\beta$ simultaneously. These can be obtained along the similar lines as discussed for the one-component elementary chirp model in subsection \ref{onecomplse}.\\
	
	\noindent For ready reference, we also present the CP function for the model \eqref{onecompmodel} as follows:\begin{equation}\label{cpf}
		CPF\left(t, \Omega \right) =\sum_{m=0}^{\frac{\left( N-1\right) }{2}  } y\left(t+m \right) y\left(t-m \right) e^{-i\Omega m^{2}}.
	\end{equation}   For the computation of the LSEs and the ALSEs, the periodogram-type estimates have been used as the initial values as discussed in section \ref{onecomplse}. The initial value in CPF method has been obtained by maximising the CPF \eqref{cpf} over the grid points $ \dfrac{\pi k}{N^{2}}, k=1, \dots, N^{2}-1 $. In the dechirping method, we have obtained initial value for $\beta$ by minimising the objective function \eqref{objfundechirp} over the Fourier grid, $\frac{\pi k}{N-1}, k=1, 2, \dots, N-2$. After computing the initial value, we use Nelder-Mead simplex algorithm to optimize the corresponding objective function in all four estimation methods, that is, the LSEs, the ALSEs, the dechirping method and the CPF method. We use in-built function ``optim" in R software for the Nelder-Mead algorithm. Here, we have restricted our grid search among 10 number of points around the true value in all the methods to save time involved in the computation.
	
	\begin{figure}[!t]
		\includegraphics[width=\linewidth, height=8cm]{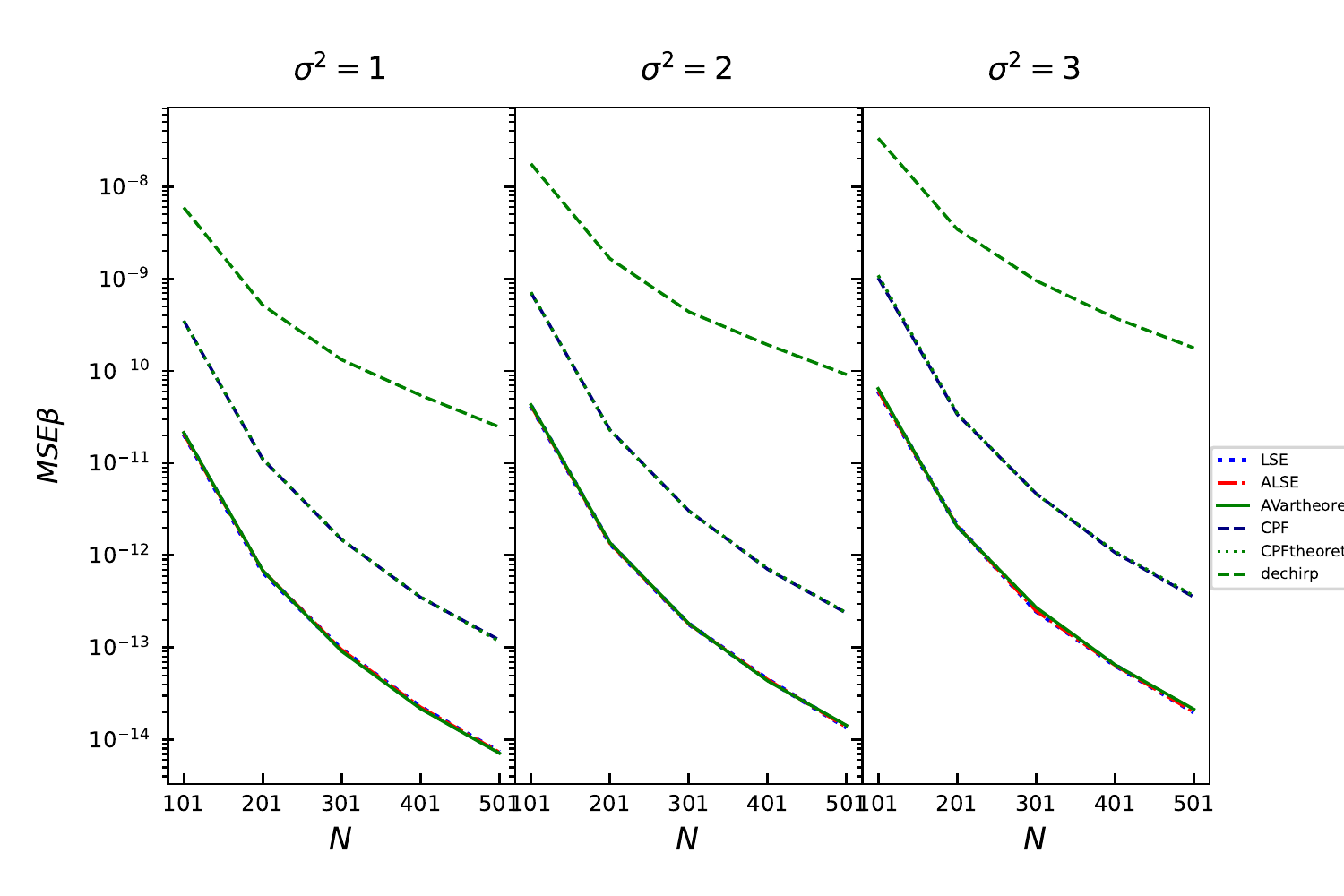}
		\caption{Mean squared errors and theoretical asymptotic variances of different estimates of frequency rate of the simulated one-component model for different error
			variances versus sample size.}
		\label{one_comp_mse}
	\end{figure}
	\begin{figure}[!t]
		\includegraphics[width=\linewidth, height=8cm]{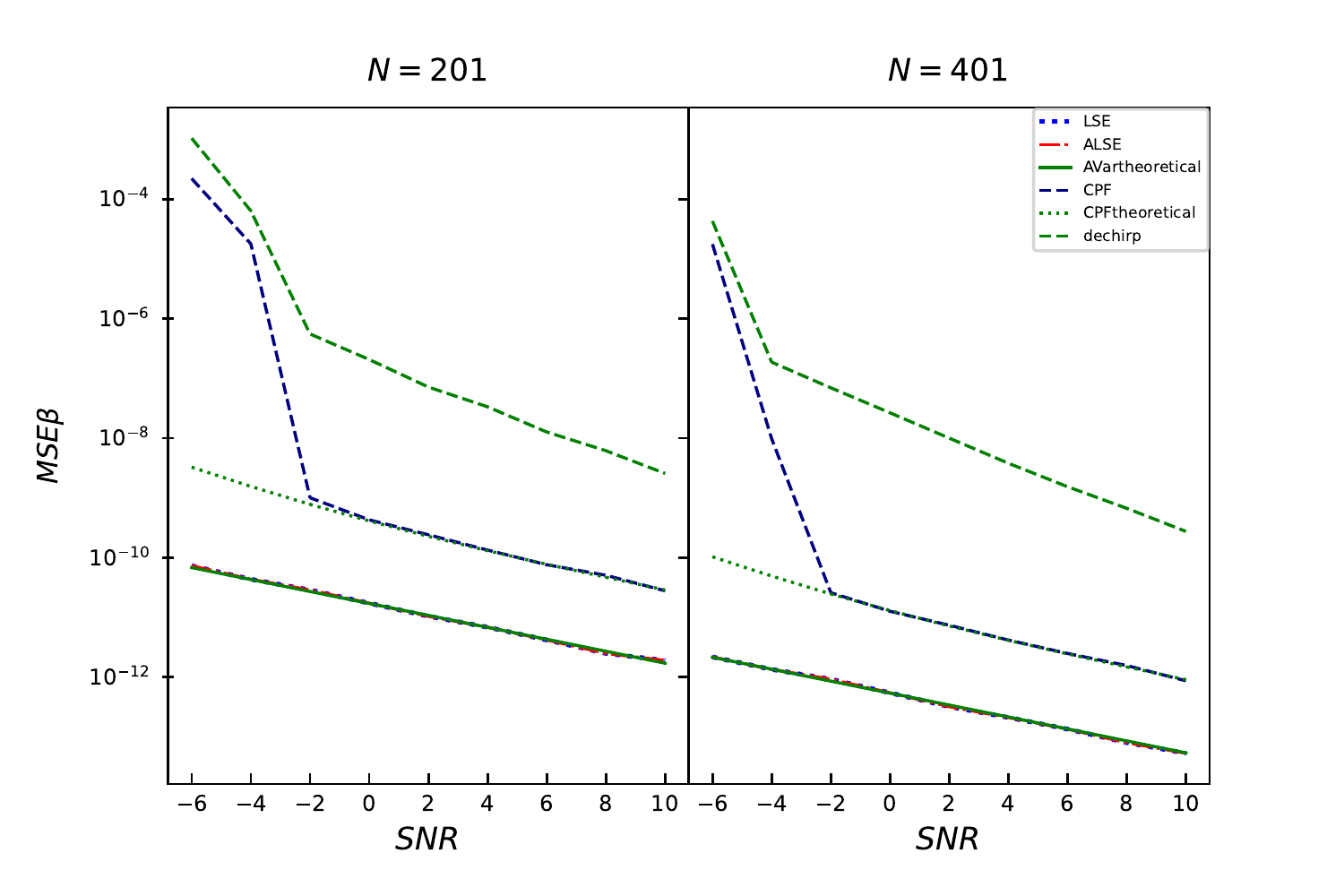}
		\caption{Mean squared errors and theoretical asymptotic variances of different estimates of frequency rate of the simulated one-component model for different sample sizes versus SNR.}
		\label{one_comp_snr}
	\end{figure}
	\begin{figure}[!t]
		\includegraphics[width=\linewidth, height=8cm]{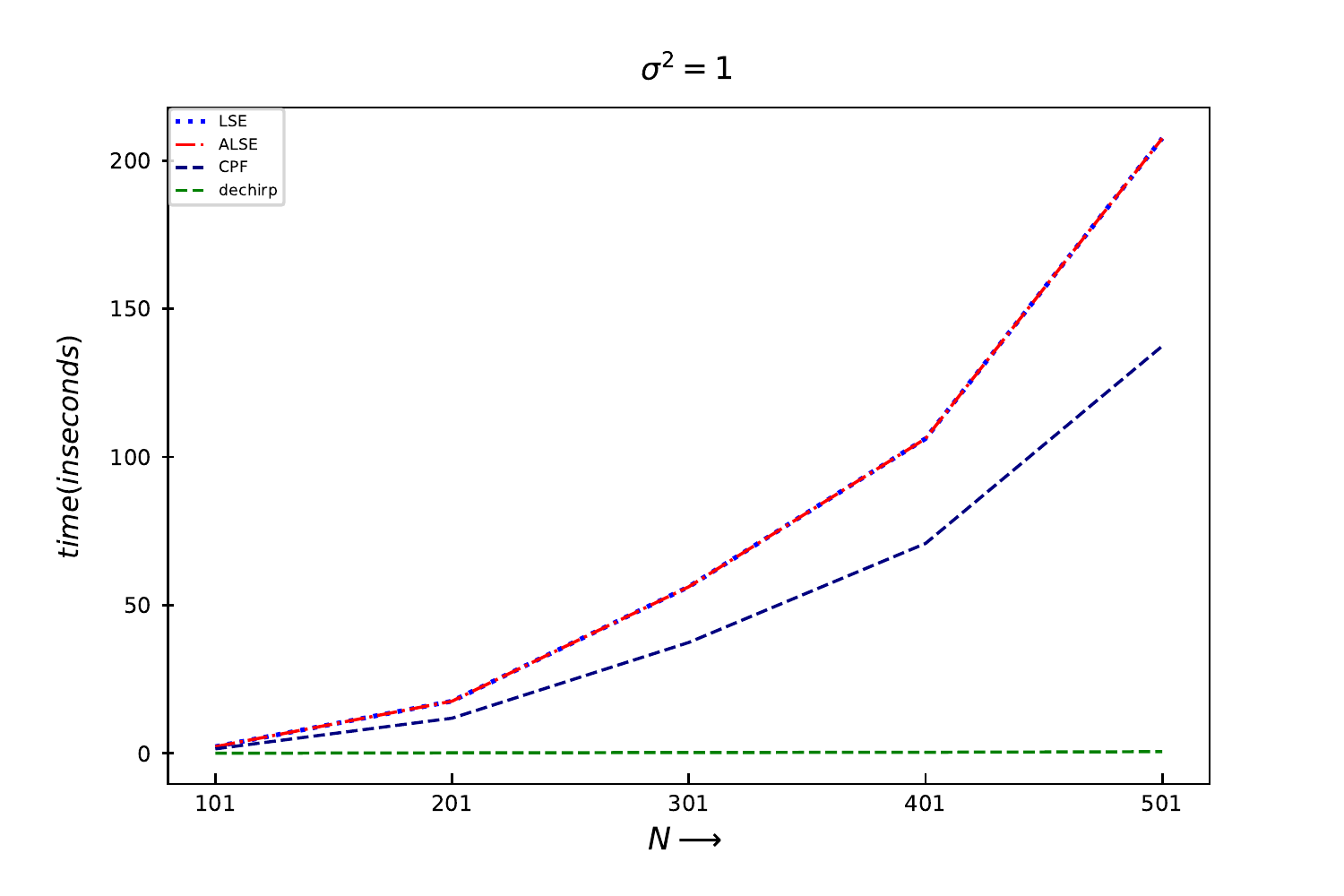}
		\caption{Computational time in estimating the frequency rate estimate using different estimation methods for the simulated one-component model versus sample size.}
		\label{one_comp_time_ev1}
	\end{figure}
	\noindent We use 1000 replications for each sample size and each error variance. We compute MSEs of frequency rate estimates using all four methods. We also calculate the theoretical asymptotic variances of the LSEs and, similarly, we calculate the theoretical asymptotic variances of the CPF method, to compare with their corresponding MSEs, we represent them as AVar theoretical and CPF theoretical, respectively, in Fig.  \ref{one_comp_mse} and Fig. \ref{one_comp_snr}.\\
	
	\noindent  MSEs and theoretical asymptotic variances of different frequency rate estimates versus different sample sizes are shown in Fig. \ref{one_comp_mse}  for different error variances. From this figure, we observe that as the sample size increases, the MSEs decreases for the LSEs as well as for the ALSEs which validates the consistency of the proposed estimators. Moreover, MSEs of the discussed estimators match pretty well with the corresponding theoretical variances for most of the cases. \\
	
	\noindent Fig. \ref{one_comp_snr} depicts MSEs of different estimators versus varying SNRs. From this figure, it is evident  that MSEs of the LSEs and the ALSEs match well with their corresponding theoretical asymptotic variances for each value of the SNR. However, for SNR below $-2$, the MSE does not match with the corresponding theoretical variance for the estimate obtained using CPF. Hence, threshold of SNR for the CPF method for this case is $-2$. This figure also shows that MSEs decreases as the SNR increases.\\
	
	\noindent In Fig.  \ref{one_comp_time_ev1}, we have plotted the time taken to compute the frequency rate estimates using the different estimation methods with respect to the diffferent sample sizes. From here, it can be observed that the  dechirping method takes the least time for estimating the frequency rate parameter as compared with the LSEs, the  ALSEs and the CPF method. LSEs and ALSEs take the maximum time in comparison to other methods but at the same time they provide the maximum statistical efficiency. Therefore, there is a trade-off between computational time and the efficiency. Further, as the sample size increases, computational time also increases for all the cases, which is quite natural. 
	
	\subsection{Simulation results for a multiple-component elementary chirp model} \label{sim2}
	\noindent We consider a two-component elementary chirp model \eqref{multicompmodel} for this simulation study with the true parameter values given as follows:
	$A_{1}^{0}=7, \beta_{1}^{0}=1, A_{2}^{0}=5$ and $\beta_{2}^{0}=0.5$.
	
	\noindent We perform these simulations for the different sample sizes and error variances same as those used for the one-component elementary chirp model simulation study. We estimate the frequency rate parameters using the sequential LSEs, the sequential ALSEs, the dechirping method and the PCPF method. For the dechirping method for multiple-component model, we first estimate the first chirp component using the LSEs as discussed in the above subsection. Then, we use sequential procedure as discussed in subsection \ref{seqlsesec} and estimate the subsequent chirp components.\\
	
	\noindent Next, we present the PCP function for the model \eqref{multicompmodel} for the given $L$ different time points, as follows:\begin{equation}\label{pcpf}
		PCPF\left(\Omega \right) =\prod_{l=1}^{L} CPF\left( t_{l}, \Omega\right).
	\end{equation} 
	Here, we have chosen two different time points, $t_{1}=0.4N$ and $t_{2}=\frac{N+1}{2}$ as in \cite{pcpf}. We have used Nelder-Mead simplex algorithm to optimize the corresponding objective function in all four estimation methods, that is, the sequential LSEs, the sequential ALSEs, the dechirping method and the PCPF method. For finding the initial values, we have done the grid search as discussed in the previous subsection \ref{sim1} for all the methods. 
	\begin{figure}[!t]
		\includegraphics[width=\linewidth, height=9cm]{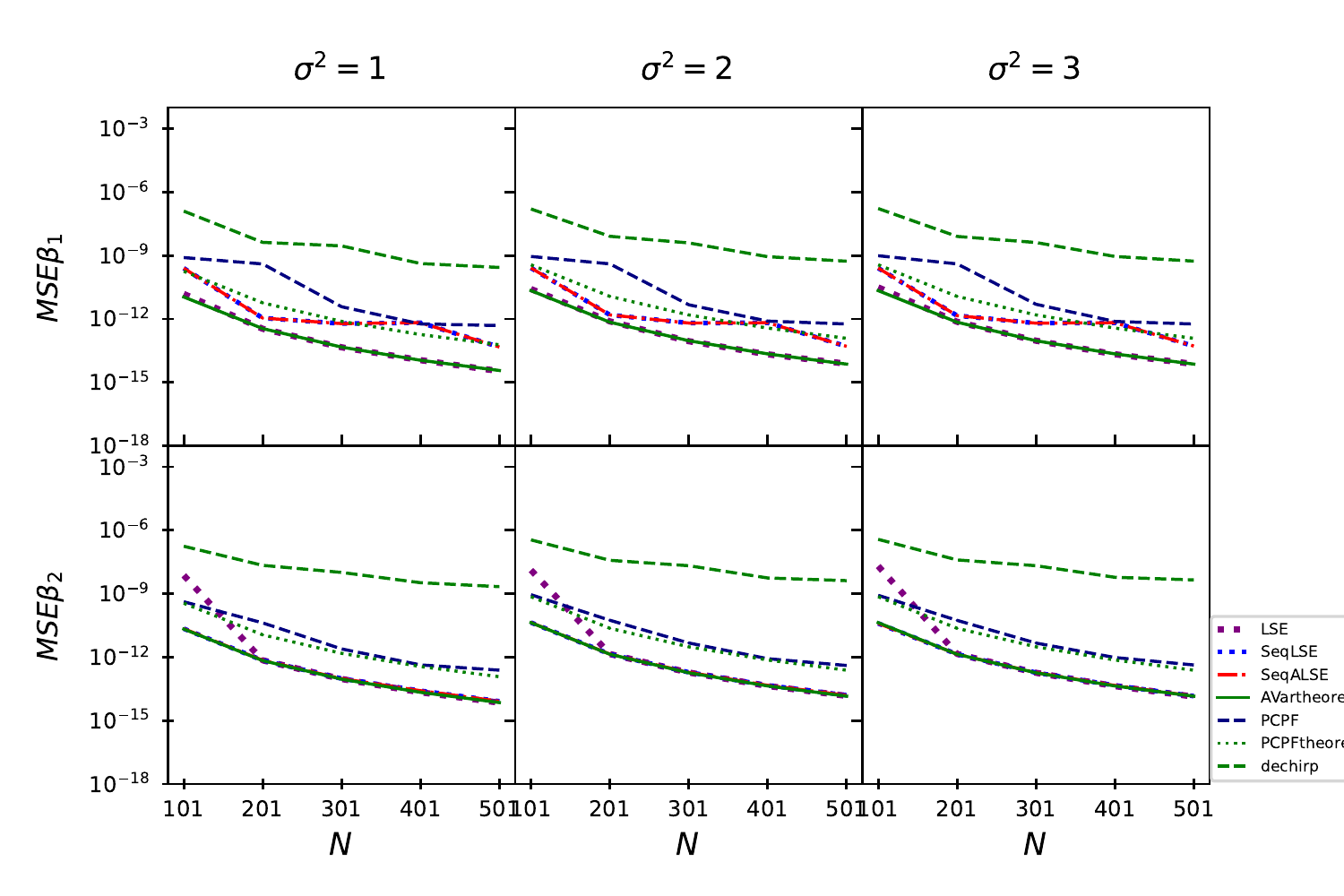}
		\caption{Mean squared errors and theoretical asymptotic variances of different estimates of frequency rates of the simulated two-component model for different error
			variances versus sample size.}
		\label{two_comp_mse}
	\end{figure}
	\begin{figure}[!t]
		\includegraphics[width=\linewidth, height=9cm]{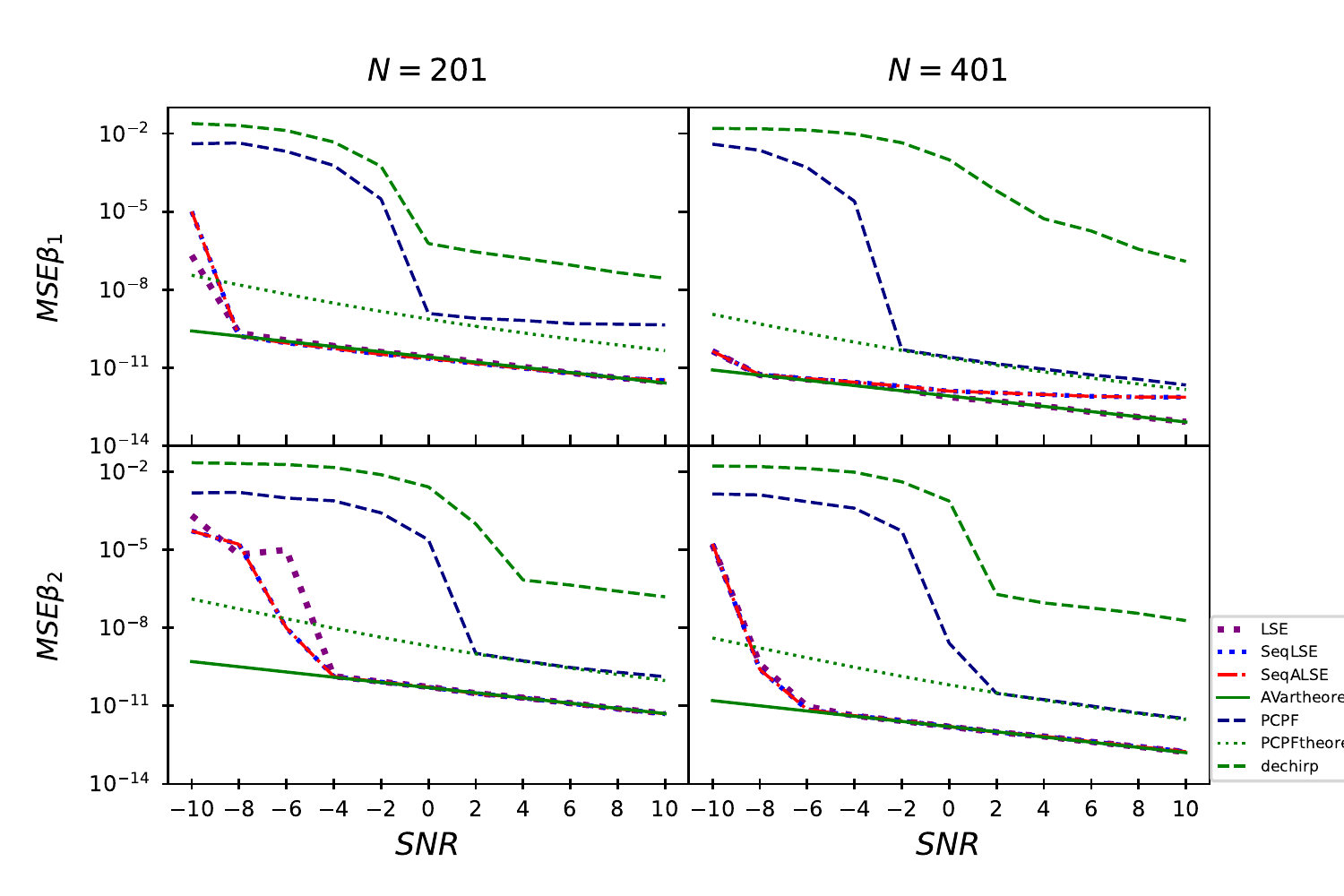}
		\caption{Mean squared errors and theoretical asymptotic variances of different estimates of frequency rates of the simulated two-component model for different sample sizes versus SNR.}
		\label{two_comp_snr}
	\end{figure}
	\begin{figure}[!t]
		\includegraphics[width=\linewidth, height=8cm]{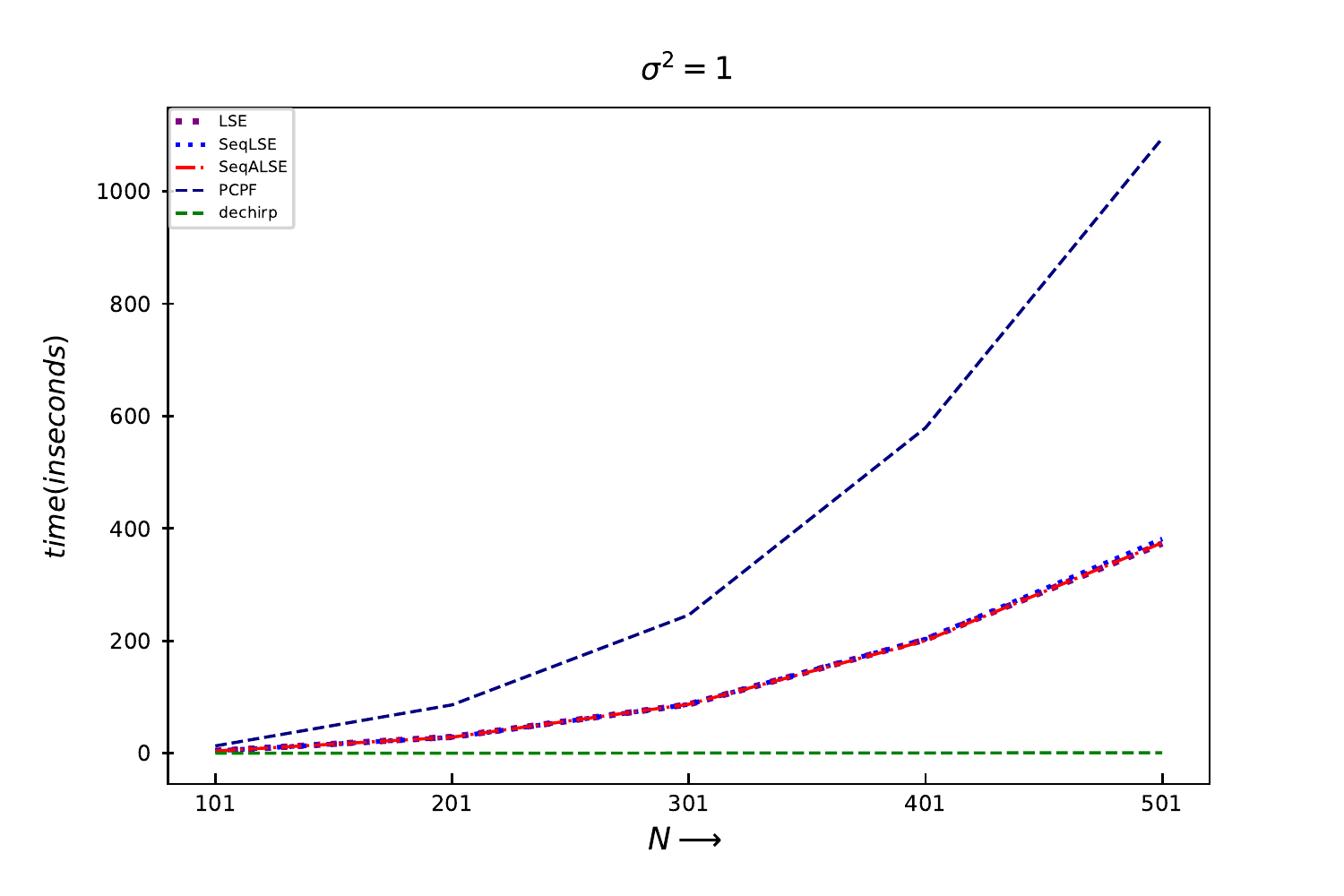}
		\caption{Computational time in estimating the different estimates of the frequency rates
			of the simulated two-component model versus sample size.}
		\label{two_comp_time_ev1}
	\end{figure} We evaluate the MSEs of the frequency rate estimates based on 1000 replications. We also calculate theoretical asymptotic variances of all the methods to compare with their corresponding MSEs. The MSEs and the theoretical asymptotic variances of different frequency rate estimates for different error variances versus different sample sizes  are shown in Fig. \ref{two_comp_mse}. In this figure, AVar theoretical and PCPF theoretical represent the theoretical asymptotic variances of the sequential  algorithm and the PCPF method, respectively. From this figure, it can also be observed that the MSEs decreases as $ N $ increases for all the methods which depicts that the frequency rate estimates get closer to the true parameter values as sample size increases. MSEs of the LSEs, the sequential LSEs, the sequential ALSEs and the estimators obtained using PCPF method match well with the corresponding theoretical asymptotic variances for most of the cases. Also, MSEs of sequential LSEs and sequential ALSEs are at par.\\
	
	\noindent Fig. \ref{two_comp_snr} shows the plot of the MSEs versus the SNR. From here, it is clear that MSEs of the LSEs of the $\beta_{2}$ match well with their corresponding theoretical asymptotic variances till the SNR value -4 when sample size is $ 201 $, and for the sequential LSEs and the sequential ALSEs, the threshold for SNR is -6. We thus observe that LSEs start performing poorly at higher SNR value than the proposed sequential estimators.  For SNR lower from $0$ and $-2$, the MSEs do not match with the corresponding theoretical variances for the estimate obtained using PCPF method, for the sample sizes $ 201  $ and $ 401 $, respectively. We also observe that MSEs decreases as the SNR increases.\\
	
	\noindent In Fig. \ref{two_comp_time_ev1}, the computational time to find the frequency rates using the different estimation methods with respect to the different sample sizes has been shown. Here, we report the time involved in computing both the frequency rates of the model. From this figure, it can be observed that the dechirping method takes the least time in comparison with the other discussed methods in this paper, not surprisingly,
	since here we have to do grid search among $N$ number of points, whereas, in other estimation methods, we have to do a grid search among $N^2$ grid points. PCPF method takes the maximum time in computation and the proposed estimators in this paper are taking more or less the same time in computing the estimates. As we have seen that the proposed estimation methods provide the estimators with the optimal rates of convergence, we recommend these estimation methods for analyzing real life data.
	
	\subsection{Simulation results for a two-component elementary chirp model when two frequency rates are close to each other}
\begin{figure}[!t]
		\includegraphics[width=\linewidth, height=8cm]{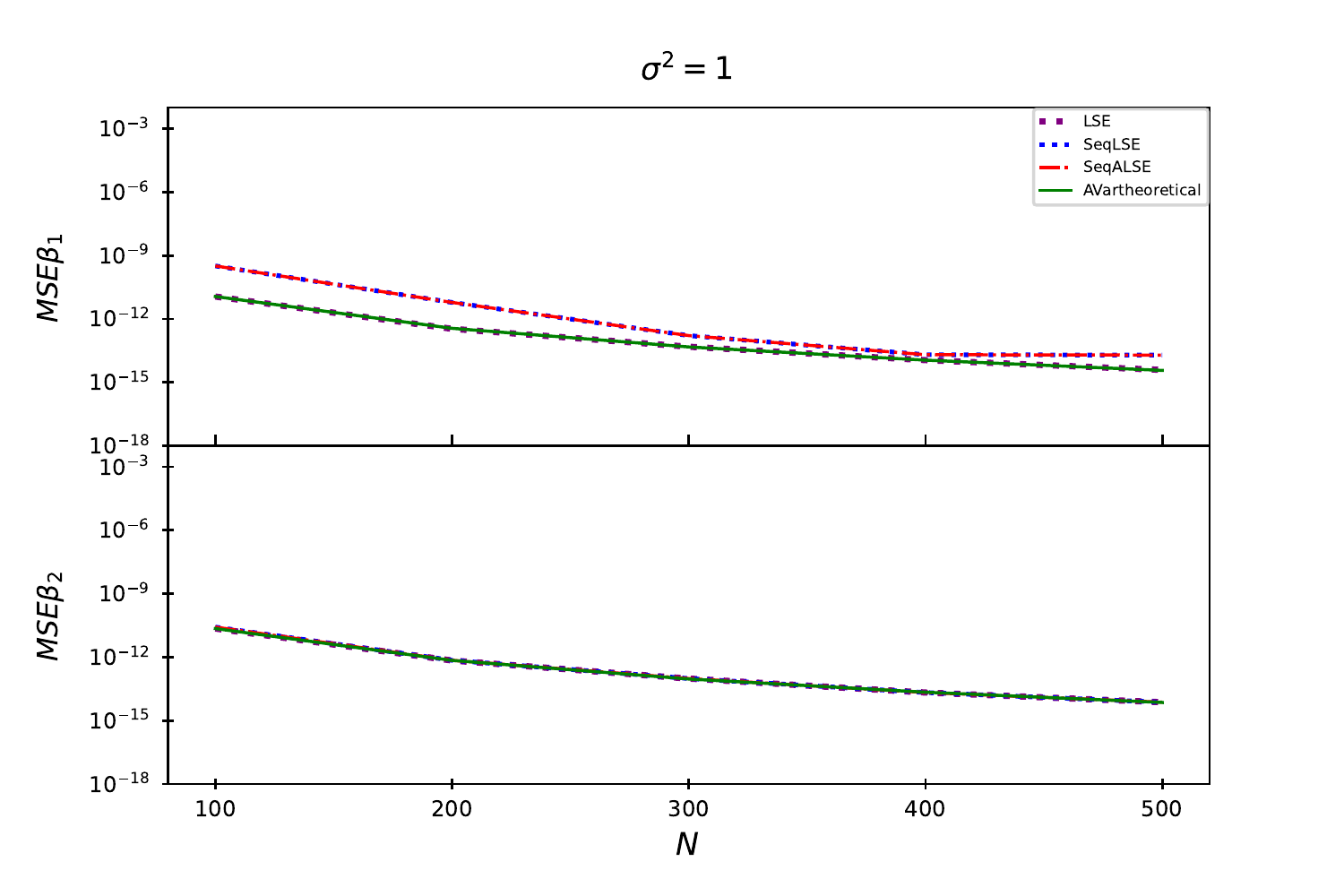}
		\caption{Mean squared errors and theoretical asymptotic variances of different estimates of frequency rates versus sample size.}
		\label{2qcb_mse}
	\end{figure}

\begin{figure}[!t]
		\includegraphics[width=\linewidth, height=8cm]{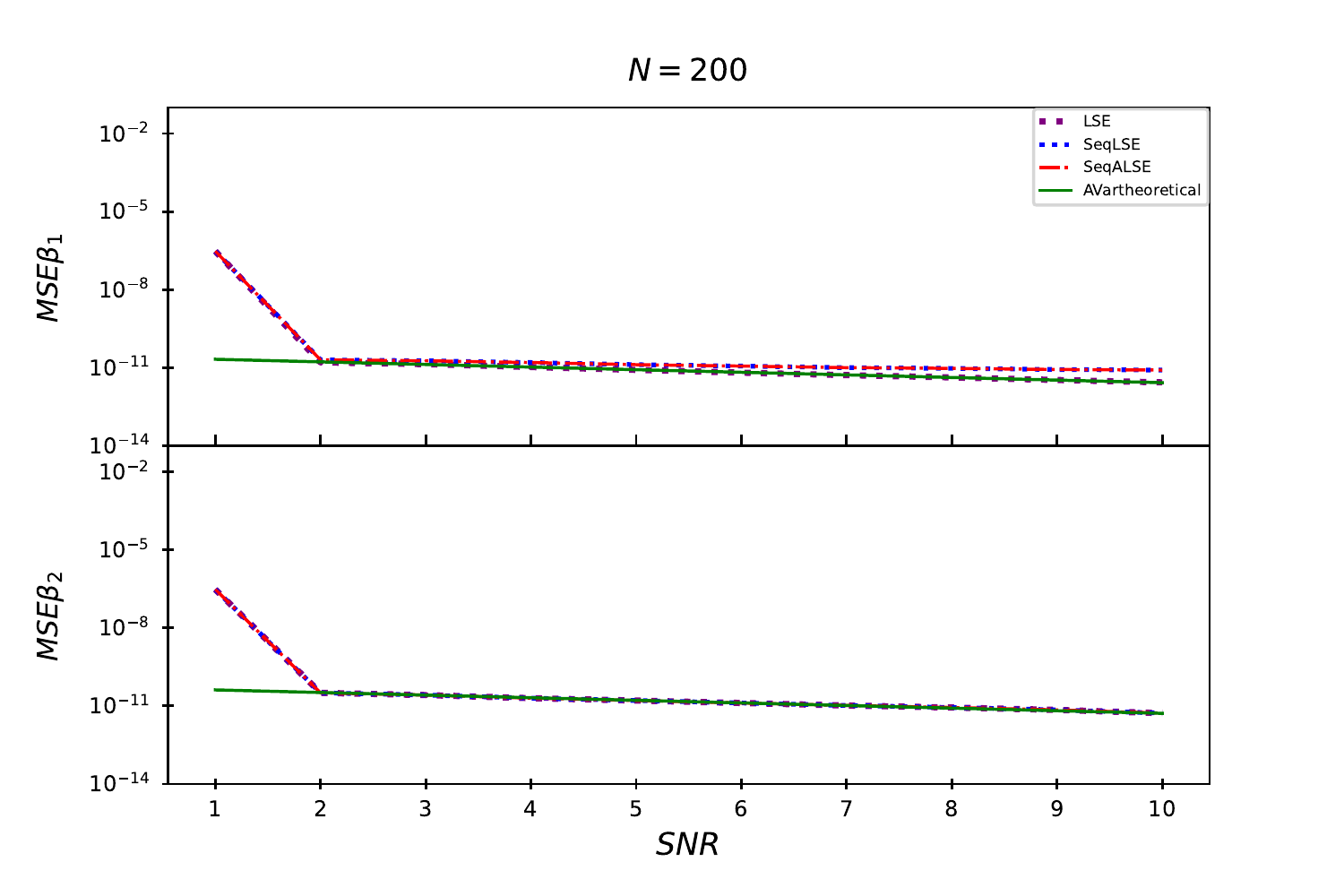}
		\caption{Mean squared errors and theoretical asymptotic variances of different estimates of frequency rates versus SNR.}
		\label{2qcb_snr}
	\end{figure}

	\noindent In this subsection, we consider a two-component elementary chirp model \eqref{multicompmodel} when two frequency rates are close to each other, to evaluate the performance of the proposed estimators under such a scenario. Here, we take different setups by fixing amplitude parameters and by bringing two frequency rates closer and closer. First, we take the gap between two frequency rates as $10^{-2}$, with the true parameter values given as: $A_{1}^{0}=7 , \beta_{1}^0=0.51, A_{2}^{0}=5 ~~ \text{and} ~~ \beta_{2}^0=0.5 $. Here, $\epsilon\left(t \right)^\prime$s are \textit{i.i.d.} complex-valued normal random variables and satisfy assumption \ref{ass1}. We take $N=100, 200, 300, 400, 500$ and $\sigma^{2}=1$. We simulate the experiment 1000 times and compute the MSEs of the frequency rate estimates. Fig. \ref{2qcb_mse} shows the plot for the MSEs versus sample sizes. From Fig. \ref{2qcb_mse}, it can be observed that MSEs obtained using the proposed estimation methods are at par with the theoretical asymptotic variances. Fig. \ref{2qcb_snr} shows the plot for MSEs versus SNR, when sample size is $200$. It can be seen that MSEs of frequency rate estimates match nicely with theoretical asymptotic variances when the SNR value is greater than 1. From these plots it is evident that two frequency rates are resolvable in this case.
	
\begin{figure}[!t]
		\includegraphics[width=\linewidth, height=8cm]{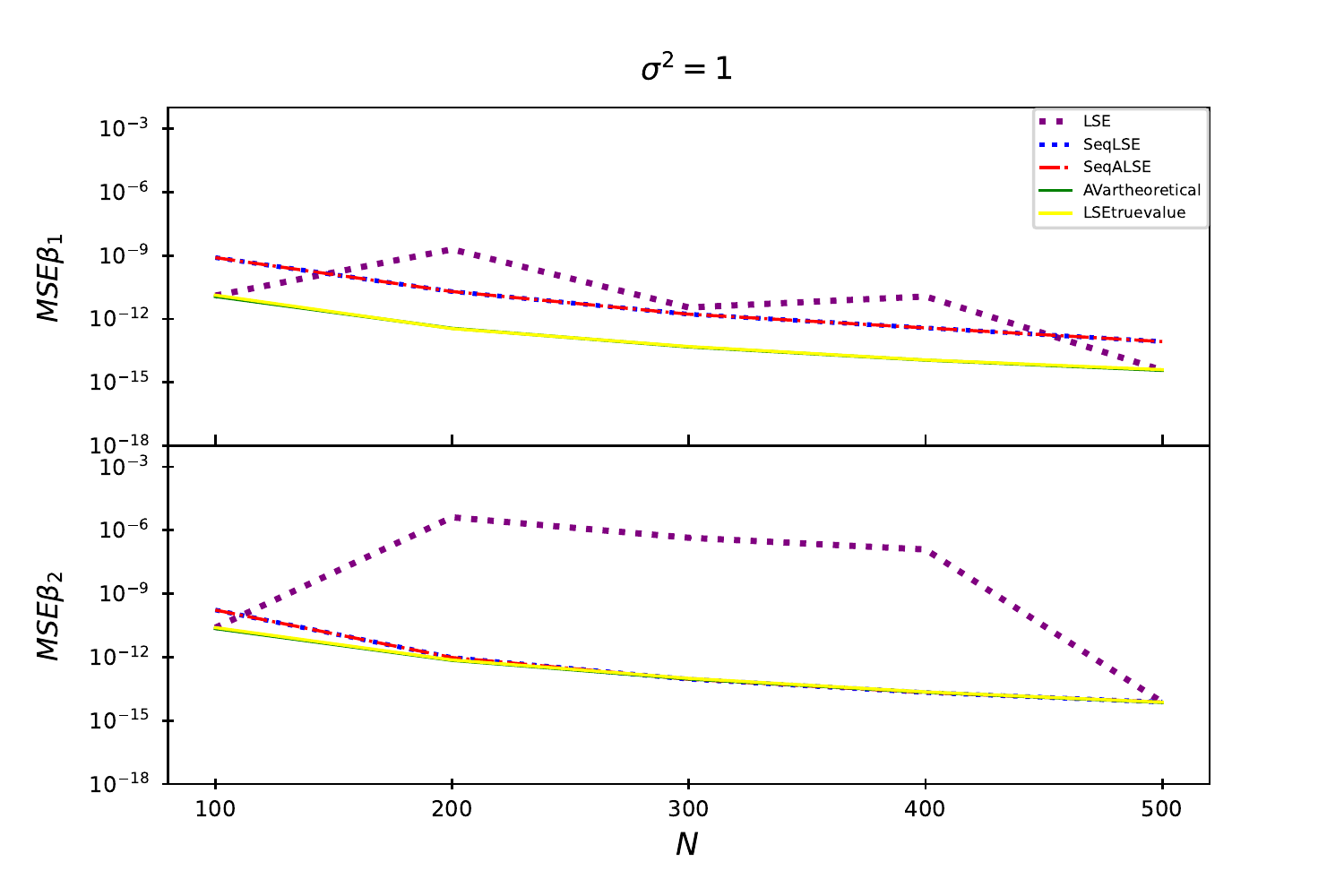}
		\caption{Mean squared errors and theoretical asymptotic variances of different estimates of frequency rates versus sample size.}
		\label{2qcb2_mse}
	\end{figure}

\begin{figure}[!t]
		\includegraphics[width=\linewidth, height=8cm]{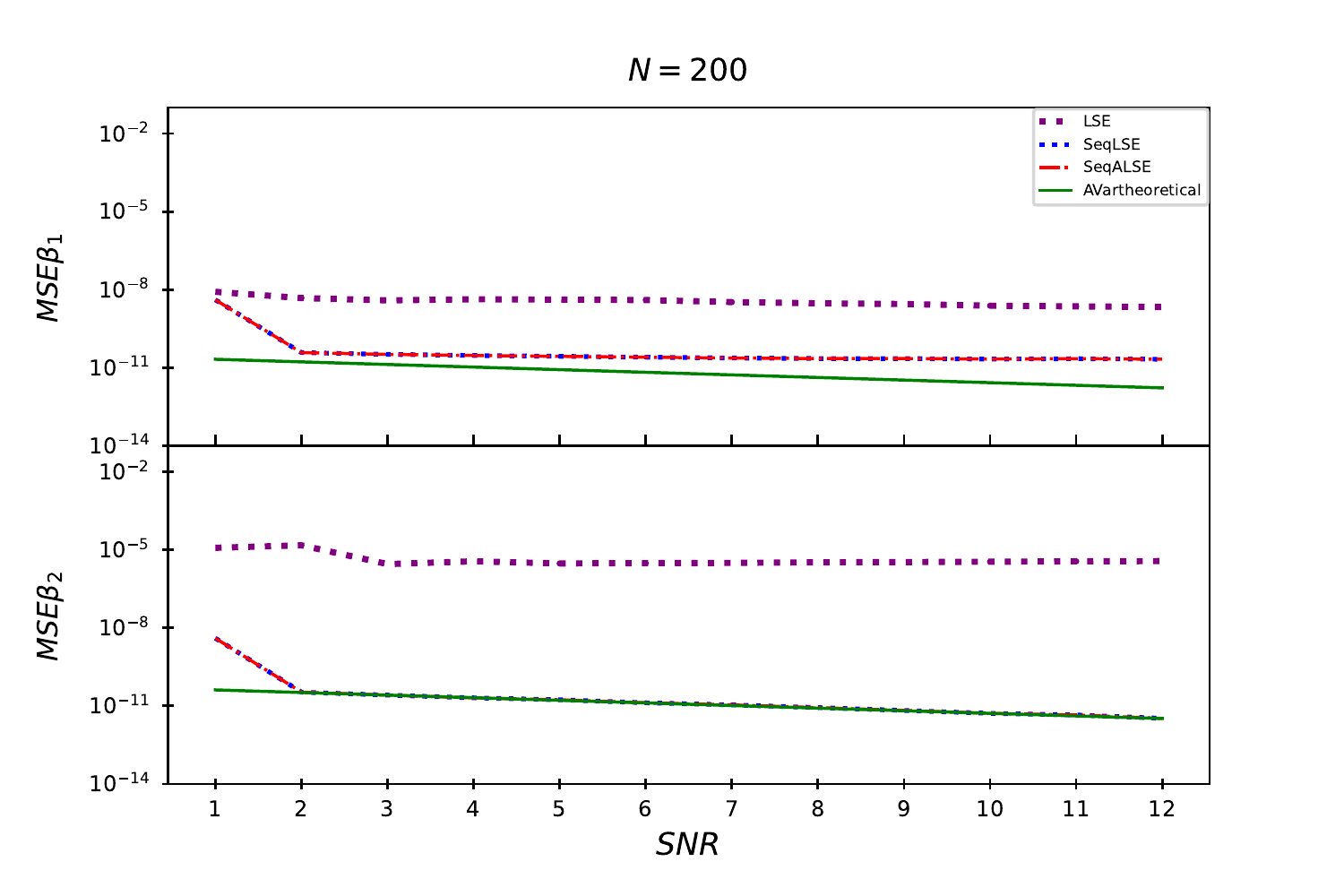}
		\caption{Mean squared errors and theoretical asymptotic variances of different estimates of frequency rates versus SNR.}
		\label{2qcb2_snr}
	\end{figure}

	\noindent We next consider the following setup: $A_{1}^{0}=7, \beta_{1}^0=0.502, A_{2}^{0}=5 ~ \text{and} ~ \beta_{2}^0=0.5 $. Note that, here the gap between two frequency rates is $2 \times 10^{-3}$. Fig. \ref{2qcb2_mse} depicts the plot for MSEs versus sample size. From this figure, it can be observed that the MSEs obtained using sequential LSEs as well as sequential ALSEs are at par with theoretical asymptotic variances. In Fig. \ref{2qcb2_mse}, LSEtruevalue is representing the MSEs of the LSEs when true parameter values are taken as the initial values. It is interesting to note that the MSEs obtained using LSEs fail to match the theoretical asymptotic variances when the initial values are obtained using the periodogram-type function. Fig. \ref{2qcb2_snr} shows the plot for MSEs versus SNR, when the sample size is $200$. MSEs of frequency rate estimates match nicely with theoretical asymptotic variances when the SNR value is greater than 1, when estimation is done using the sequential LSEs and the sequential ALSEs. In case of the LSEs, they do not match with the theoretical asymptotic variances. From these results we observe that although the LSEs are not stable in this case but, the sequential LSEs and the sequential ALSEs give satisfactory and stable results.\\
	
\begin{figure}[!t]
		\includegraphics[width=\linewidth, height=8cm]{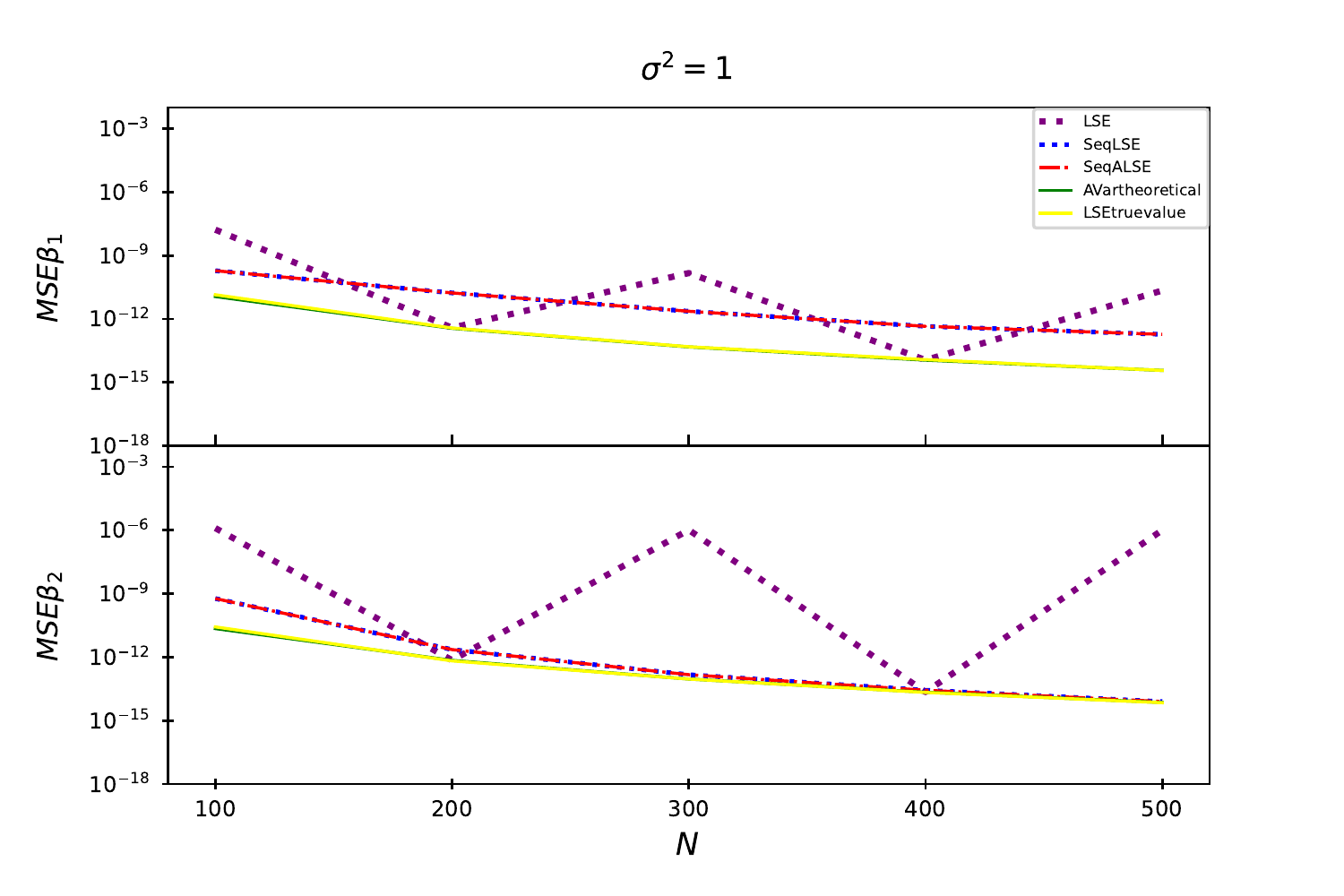}
		\caption{Mean squared errors and theoretical asymptotic variances of different estimates of frequency rates versus sample size.}
		\label{2qcb3_mse}
	\end{figure}

\begin{figure}[!t]
		\includegraphics[width=\linewidth, height=8cm]{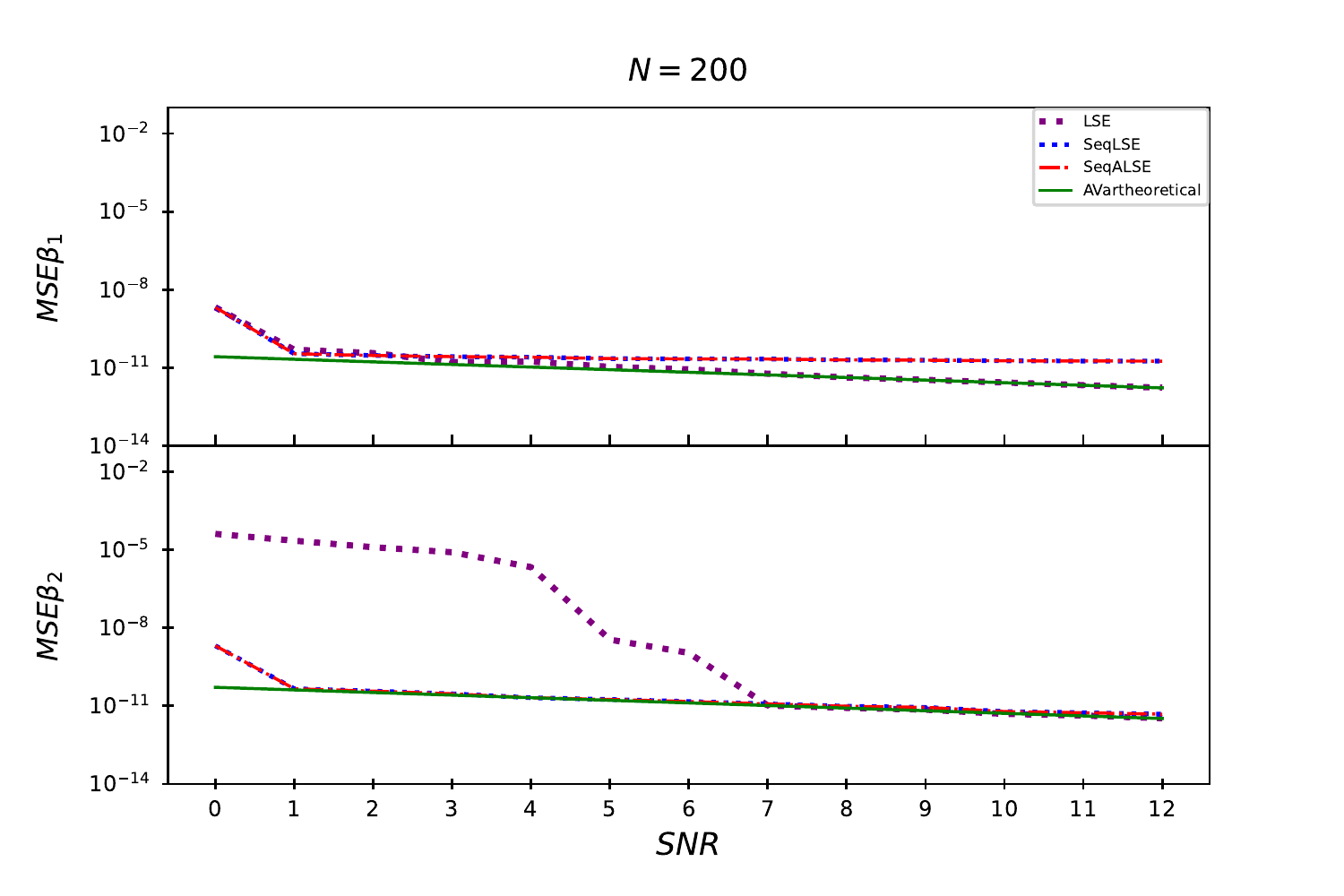}
		\caption{Mean squared errors and theoretical asymptotic variances of different estimates of frequency rates versus SNR.}
		\label{2qcb3_snr}
	\end{figure}

	\noindent Now, we take $A_{1}^{0}=7, \beta_{1}^0=0.501, A_{2}^{0}=5 ~ \text{and} ~ \beta_{2}^0=0.5 $. Note that, here the gap between two frequency rates is $ 10^{-3}$. Fig. \ref{2qcb3_mse} represents the plot for MSEs versus sample sizes. From this plot, it can be observed that the MSEs obtained using sequential LSEs as well as sequential ALSEs are at par with the theoretical asymptotic variances. In Fig. \ref{2qcb3_mse}, LSEtruevalue is representing the MSEs of the LSEs when true parameter values are taken as the initial values. Here, it can be observed that the MSEs obtained using LSEs fail to match the theoretical asymptotic variances when the initial values are obtained using the periodogram-type function. Fig. \ref{2qcb3_snr} gives the plot for MSEs versus SNR, when the sample size is $200$. MSEs of frequency rate estimates match nicely with theoretical asymptotic variances when the SNR is greater than 0, when estimation is done using sequential LSEs and sequential ALSEs. In case of the LSEs, threshold of the SNR is 7, which is quite high SNR threshold as compared to the threshold of the sequential estimators. From these results, we can conclude that the two close frequency rates are not resolvable after a certain threshold using the LSEs, however sequential LSEs and sequential ALSEs are able to resolve them.\\
	
\begin{figure}[!t]
		\includegraphics[width=\linewidth, height=8cm]{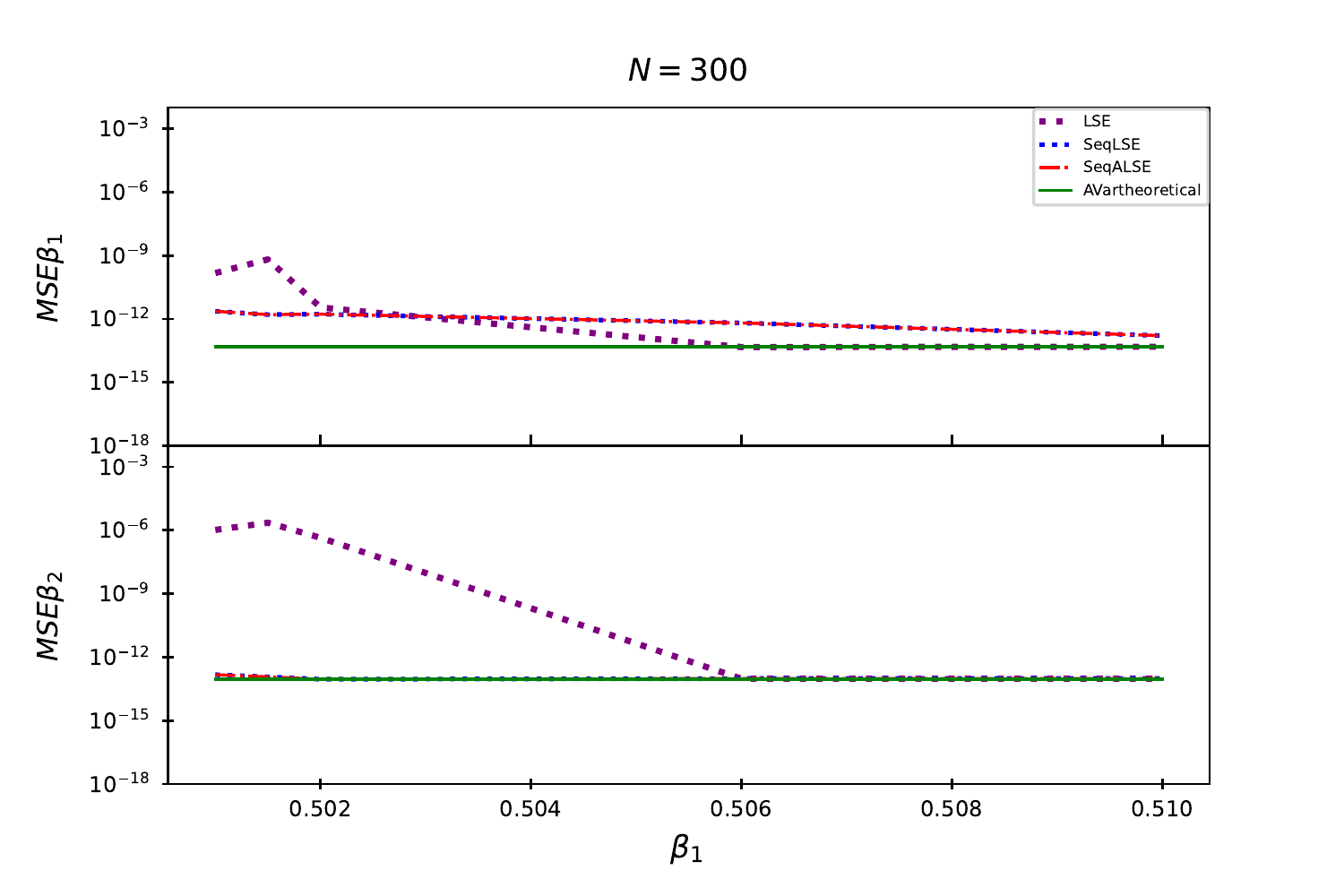}
		\caption{Mean squared errors and theoretical asymptotic variances of different estimates of frequency rates versus frequency rate. }
		\label{dbet_mse}
	\end{figure}

	\noindent In the next set of simulations, we fix $\beta_{2}$ at $0.5$ and vary $\beta_{1}$, taking it close to $\beta_{2}$. Values of $\beta_{1}$ are taken as $0.501, 0.5015, 0.502, 0.506, 0.51$. The values of the amplitude parameters are $A_{1}^{0}=7 ~ \text{and} ~ A_{2}^{0}=5$, sample size is taken as $N=300$ and error variance is $\sigma^{2}=1$.
	Fig. \ref{dbet_mse} shows the plot for MSEs versus $\beta_{1}$. We observe that the MSEs obtained using LSEs do not match well with the theoretical asymptotic variances when $\beta_{1}$ gets closer to $\beta_{2}$. MSEs obtained using sequential LSEs and sequential ALSEs are close to the theoretical asymptotic variance.
	
	\section{Real Data Analysis} \label{realdatana}
	\begin{figure}[!t]
		\includegraphics[width=\linewidth, height=8cm]{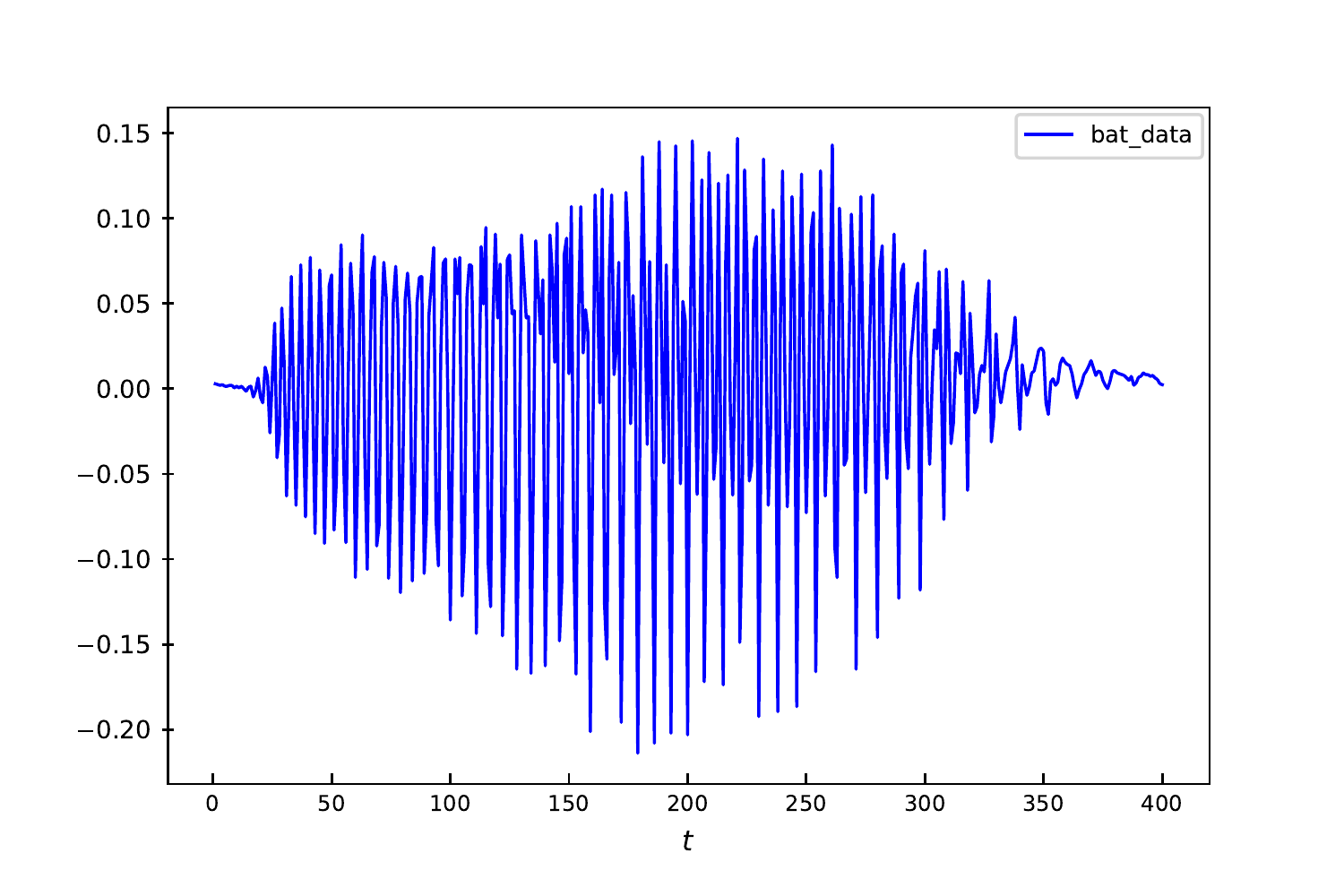}
		\caption{Bat data}
		\label{batdatoriginal}
	\end{figure}
	\begin{figure}[!t]
		\includegraphics[width=\linewidth, height=8cm]{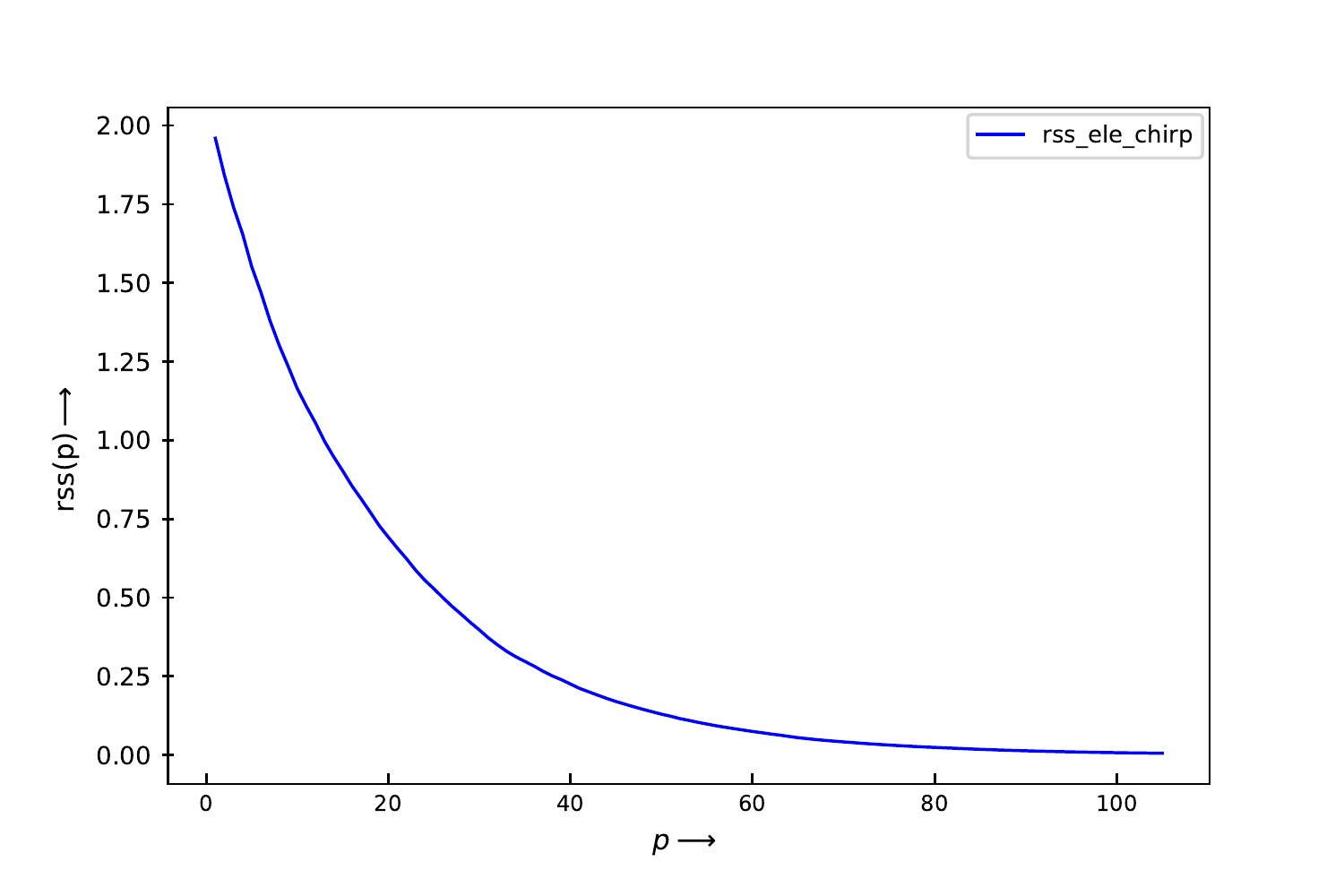}
		\caption{Residuals sum of squares for bat data using sequential LSE.}
		\label{rss_plot}
	\end{figure}
	\begin{figure}[!t]
		\includegraphics[width=\linewidth, height=8cm]{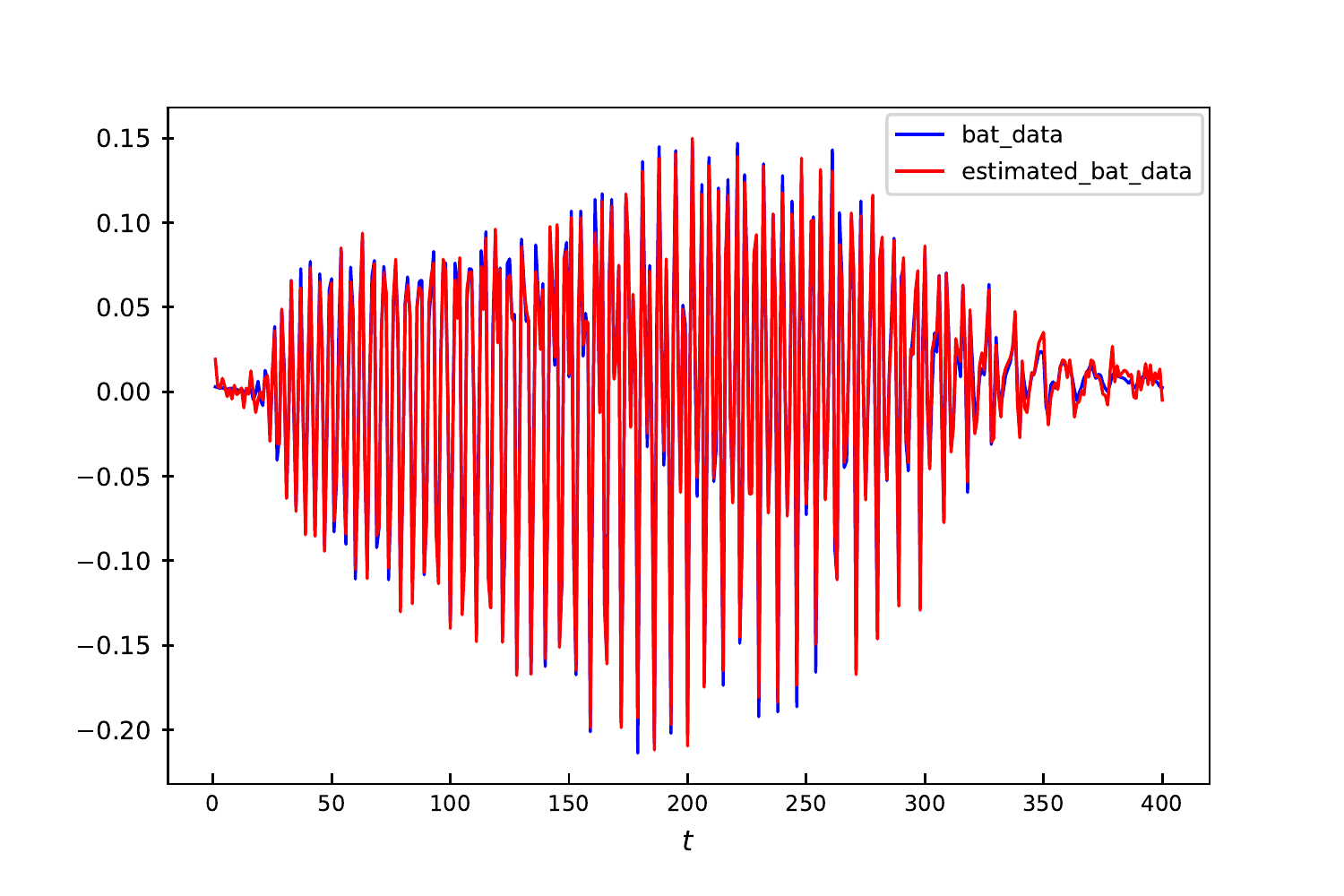}
		\caption{Observed bat signal and estimated bat signal using elementary chirp model. }
		\label{est_bata_dat_ele_chirp}
	\end{figure}
	\begin{figure}[!t]
		\includegraphics[width=\linewidth, height=8cm]{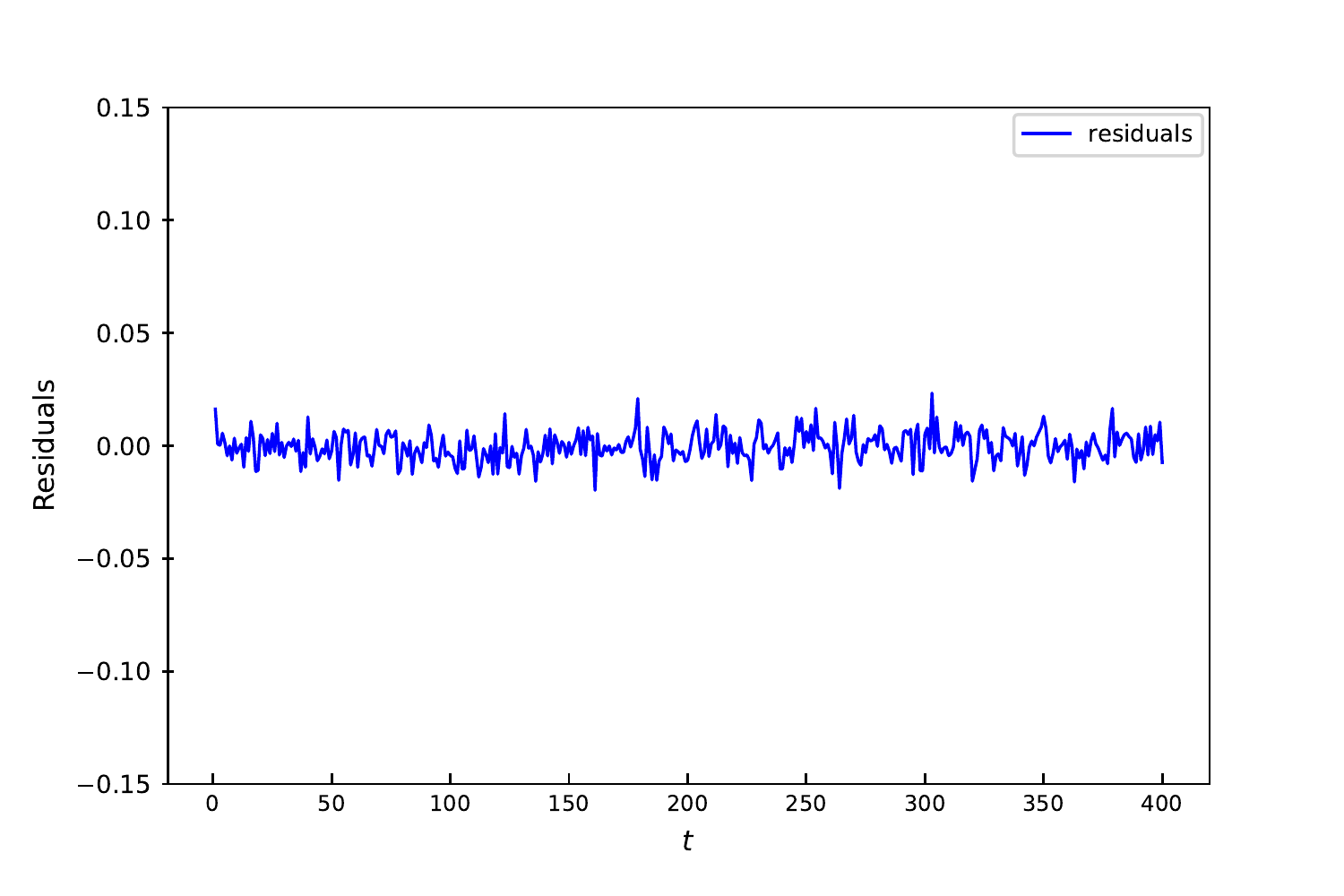}
		\caption{Residuals at 85th elementary chirp component.}
		\label{res85elechirp}
	\end{figure}
	
	\noindent In this section, we demonstrate a real data analysis which shows the applicability of the elementary chirp model in real world. Here, we consider a data of ultrasound produced by bat \cite{alsechirplike}. The original signal can be seen in Fig. \ref{batdatoriginal}, which is of length 400. Note that, we have seen that the proposed sequential estimators provide the optimal estimators with lower computational complexity. Therefore, we fit an elementary chirp model to this data using the proposed sequential LSE method. We observe that the traditional information theoretic criterion does not work nicely for this data. Therefore, we fit it at the component after which residual sum of squares (RSS) does not decrease much in comparison with the previous step RSS. We take a maximum of 105 elementary chirp components to fit the data and fit it at the 85th component due to the above discussed reason which is evident in Fig. \ref{rss_plot}. In Fig. \ref{est_bata_dat_ele_chirp}, we plot the fitted data along with the original signal, using 85 elementary chirp components. The figure indicates that the fit of the data using the proposed sequential LSE is quite satisfactory and the fit matches the observed data quite well.
	
	\noindent We have also performed residual analysis to test the error assumption of the model. We use Ljung-Box test \cite{ljungbox} for this purpose. We use the in-built function “Box.test” in R software. For this data set, the Ljung-Box test does not reject the null hypothesis. Therefore, by using Ljung-Box test and looking into the residuals plot in Fig. \ref{res85elechirp}, we may conclude that the residuals
	are \textit{i.i.d.}. Thus, the elementary chirp model fits well the considered data using the proposed sequential estimation method.

	\section{Conclusion} \label{concludesec}
	\noindent In this paper, we propose some parameter estimation methods to estimate the elementary chirp model parameters. It has been observed that the proposed estimators, namely, LSEs, ALSEs, sequential LSEs, and sequential ALSEs provide estimators of the unknown parameters with the optimal rates of convergence. From the simulation studies, it can be seen that the proposed estimators work well because their MSEs match the theoretical asymptotic variances. Further, the proposed sequential estimators provide the estimators of the unknown parameters with the same optimal theoretical properties as the LSEs and  ALSEs with lower computational complexity. It is also observed that the sequential LSEs and sequential ALSEs  are able to resolve the frequency rates when two frequency rates are close to each other, while LSEs are unstable in such a case. Therefore, we recommend the proposed sequential estimators for real signal analysis. A real data analysis has been performed using the proposed sequential LSEs. It illustrates the practical importance of the model and assessment of the proposed sequential estimators.

	\section*{Appendix A} \label{onecompapp}
	
	\subsection*{A.1}\textbf{Proofs of the theoretical properties of the LSEs for one-component model}  \label{lseonecompapp}
	\begin{lemma} \label{consle}
		Consider \(S_{c} = \left\lbrace \bm{\theta}: \left|\bm{\theta}-\bm{\theta}^{0} \right| > c; \bm{\theta} \in \bm{\Theta}_{1} \right\rbrace\). If for any $  c>0 $,
		\begin{equation} \label{coneq}
			\operatorname{lim} \operatorname{inf} \underset{\bm{\theta} \in S_{c}}{\operatorname{inf}} \frac{1}{N} \left[Q\left(\bm{\theta} \right) - Q\left( \bm{\theta}^{0}\right)   \right] >0 ~~  a.s.,
		\end{equation}
		then $\hat{\bm{\theta}} \xrightarrow{a.s.} \bm{\theta}^{0}$ as $N \rightarrow \infty$.
	\end{lemma}
	\textit{Proof:}	Proof of this lemma can be obtained in the similar manner of lemma 5 of \cite{lsechirplike}.

	\noindent \textit{Proof of Theorem \ref{consthonecomplse}}: Note that we can write
	\begin{equation*}
		\frac{1}{N} \left[Q\left(\bm{\theta} \right) - Q\left( \bm{\theta}^{0}\right)   \right] = f\left( \bm{\theta}\right) +g\left( \bm{\theta}\right) ,	
	\end{equation*}
	where
	\begin{equation*}
	\begin{split}
		f\left( \bm{\theta}\right)=&\frac{1}{N} \sum_{t=1}^{N} \left\lbrace \left(A_{R}^{0} \cos\left( \beta^{0} t^{2}\right) - A_{I}^{0}\sin\left(\beta^{0} t^{2}\right) - A_{R} \cos\left( \beta t^{2}\right) + A_{I}\sin\left(\beta t^{2}\right) \right)^{2} \right. \\ & \left.+ \left(A_{I}^{0} \cos\left( \beta^{0} t^{2}\right) + A_{R}^{0}\sin\left(\beta^{0} t^{2}\right) - A_{I} \cos\left( \beta t^{2}\right) - A_{R}\sin\left(\beta t^{2}\right) \right)^{2}\right\rbrace ,
	\end{split}
\end{equation*}and
\begin{equation*}
	\begin{split}
		g\left( \bm{\theta}\right)=&\frac{2}{N} \sum_{t=1}^{N} \left\lbrace \epsilon_{R}\left(t \right) \left(A_{R}^{0} \cos\left( \beta^{0} t^{2}\right) - A_{I}^{0}\sin\left(\beta^{0} t^{2}\right) - A_{R} \cos\left( \beta t^{2}\right) + A_{I}\sin\left(\beta t^{2}\right) \right) \right. \\ & \left.+\epsilon_{I}\left(t \right) \left(A_{I}^{0} \cos\left( \beta^{0} t^{2}\right) + A_{R}^{0}\sin\left(\beta^{0} t^{2}\right) - A_{I} \cos\left( \beta t^{2}\right) - A_{R}\sin\left(\beta t^{2}\right) \right)\right\rbrace .
	\end{split}
\end{equation*}
	
	Now using lemma 4 of \cite{lsechirplike}, it can be proved that:
	\begin{equation}
		\underset{N\rightarrow \infty}{\operatorname{lim}}\underset{\bm{\theta}\in S_{c}}{\operatorname{sup}} g\left(\bm{\theta} \right)  = 0 ~~~ a.s..
	\end{equation}
	Therefore, the following holds, \begin{equation*}
		\operatorname{lim} \operatorname{inf} \underset{\bm{\theta} \in S_{c}}{\operatorname{inf}} \frac{1}{N} \left[Q\left(\bm{\theta} \right) - Q\left( \bm{\theta}^{0}\right)   \right] = \operatorname{lim} \operatorname{inf} \underset{\bm{\theta} \in S_{c}}{\operatorname{inf}}f\left( \bm{\theta}\right) .
	\end{equation*}
	Consider the following set 
 $S_{c} = \left\lbrace \bm{\theta}: \left|\bm{\theta}-\bm{\theta}^{0} \right| \geq 3c; \bm{\theta} \in \bm{\Theta} \right\rbrace \subset S_{c1} \cup S_{c2} \cup S_{c3}  = S$, where,

\noindent $S_{c1}=\left\lbrace \bm{\theta}: \left| A_{R} - A_{R}^{0} \right| \geq c; \bm{\theta} \in \bm{\Theta}  \right\rbrace $, $S_{c2}=\left\lbrace \bm{\theta}: \left| A_{I} - A_{I}^{0} \right| \geq c; \bm{\theta} \in \bm{\Theta}  \right\rbrace $ and \\
$S_{c3}=\left\lbrace \bm{\theta}: \left| \beta - \beta^{0} \right| \geq c; \bm{\theta} \in \bm{\Theta}  \right\rbrace $. 
Thus,
	\begin{equation}
		\operatorname{lim} \operatorname{inf} \underset{\bm{\theta} \in S_{c}}{\operatorname{inf}}f\left( \bm{\theta}\right) \geq \operatorname{lim} \operatorname{inf} \underset{\bm{\theta} \in S}{\operatorname{inf}}f\left( \bm{\theta}\right).
	\end{equation} 
	Now, set $S_{c1}$ is divided as shown below:\\
$S_{c1}=\left\lbrace \bm{\theta}: \left| A_{R} - A_{R}^{0} \right| \geq c; \bm{\theta} \in \bm{\Theta}  \right\rbrace \subset S_{c1}^{1} \cup S_{c1}^{2}$ \\
$S_{c1} \subset \left\lbrace \bm{\theta}: \left| A_{R} - A_{R}^{0} \right| \geq c; \bm{\theta} \in \bm{\Theta}, \beta = \beta^{0}  \right\rbrace \cup \left\lbrace \bm{\theta}: \left| A_{R} - A_{R}^{0} \right| \geq c; \bm{\theta} \in \bm{\Theta}, \beta \neq \beta^{0}  \right\rbrace $.\\
	Using lemma 2 of \cite{lsechirplike}, it follows that 
	\begin{equation*}
		\operatorname{lim} \operatorname{inf} \underset{\bm{\theta} \in S_{c1}^{1}}{\operatorname{inf}}f\left( \bm{\theta}\right)>0~ a.s. ~ \text{as} ~ N \rightarrow \infty,
	\end{equation*}
	\begin{equation*}
		\operatorname{lim} \operatorname{inf} \underset{\bm{\theta} \in S_{c1}^{2}}{\operatorname{inf}}f\left( \bm{\theta}\right) >0~ a.s. ~ \text{as} ~ N \rightarrow \infty.
	\end{equation*}
	Hence, $\operatorname{lim} \operatorname{inf} \underset{\bm{\theta} \in S_{c1}}{\operatorname{inf}}f\left( \bm{\theta}\right) >0$. In the similar manner, it can be shown for the remaining sets. Thus, strong consistency  of $\hat{\bm{\theta}}$ follows from lemma \ref{consle}. \\

	\noindent \textit{Proof of Theorem \ref{asydthonecomplse}}: Using multivariate Taylor series expansion, expand $Q^{'}(\hat{\bm{\theta}} ) $ 
	around the point $\bm{\theta}^{0}$,
	\begin{equation} \label{taylorlseonecomp}
		Q^{'}(\hat{\bm{\theta}} )-Q^{'}(\bm{\theta}^{0} ) = (\hat{\bm{\theta}}-\bm{\theta}^{0}) Q^{''}(\bar{\bm{\theta}} ).
	\end{equation}
	Here, $\bar{\bm{\theta}}$ lies between $\bm{\theta}^{0}$ and $\hat{\bm{\theta}}$. Since $\hat{\bm{\theta}} $ minimizes $ Q(\bm{\theta} ) $, we have $	Q^{'}(\hat{\bm{\theta}} )=0$ and thus we rewrite \eqref{taylorlseonecomp} as:
	\begin{equation} \label{taylorlseonecomp2}
		(\hat{\bm{\theta}}-\bm{\theta}^{0})= -Q^{'}(\bm{\theta}^{0} )[ Q^{''}(\bar{\bm{\theta}} )]^{-1} .
	\end{equation}
	Now, multiply both side of \eqref{taylorlseonecomp2} by $\bm{D}^{-1}$, where $\bm{D}=diag\left(\frac{1}{\sqrt{N}}, \frac{1}{\sqrt{N}}, \frac{1}{N^{2}\sqrt{N}}\right) $, and obtain
	\begin{equation} \label{taylorlseonecomp3}
		(\hat{\bm{\theta}}-\bm{\theta}^{0})\bm{D}^{-1}= -Q^{'}(\bm{\theta}^{0} )\bm{D}[\bm{D} Q^{''}(\bar{\bm{\theta}} )\bm{D}]^{-1} .
	\end{equation}
	Further, using Lindeberg Feller CLT and lemma 2 of \cite{lsechirplike}, it can be shown that
	\begin{equation} \label{firderonecomplse}
		Q^{'}(\bm{\theta}^{0} ) \bm{D} \xrightarrow{d} \mathcal{N}_{3}\left( 0, \sigma^{2} \Sigma\right);
	\end{equation} where \begin{equation} \label{sigma}
		\Sigma=\begin{bmatrix}
			2 &0 & -\frac{2}{3}A_{I}^{0}\\
			0&2&\frac{2}{3}A_{R}^{0}\\
			-\frac{2}{3}A_{I}^{0} &\frac{2}{3}A_{R}^{0}& \frac{2}{5}\left| A^{0}\right|^{2} 
		\end{bmatrix}.
	\end{equation}
	Using the result that $\hat{\bm{\theta}}\xrightarrow{a.s.} \bm{\theta}^{0}$ as $N\rightarrow \infty$, we get
	\begin{equation*}
		\underset{N\rightarrow \infty}{\operatorname{lim}} \bm{D} Q^{''}(\bar{\bm{\theta}} )\bm{D} =	\underset{N\rightarrow \infty}{\operatorname{lim}} \bm{D} Q^{''}(\bm{\theta}^{0} )\bm{D}.
	\end{equation*}
	
	\noindent  The following can be proved by using lemmas 2 and 4 of \cite{lsechirplike},
	\begin{equation}\label{secderonecomplse}
		\underset{N\rightarrow \infty}{\operatorname{lim}} \bm{D} Q^{''}(\bm{\theta}^{0} )\bm{D}= \Sigma.
	\end{equation}
	Using \eqref{taylorlseonecomp3},  \eqref{firderonecomplse} and \eqref{secderonecomplse}, the desired result is obtained.
	
	\subsection*{A.2}\textbf{Proofs of the theoretical properties of the ALSEs for one-component model}\label{alseonecompapp}
	\begin{lemma} \label{conslealseonecomp}
		Suppose $\tilde{\beta}$ is the ALSE of $\beta^{0}$ and $S_{c}=\left\lbrace \beta: \left| \beta - \beta^{0} \right|>c \right\rbrace $. If for any $c>0$,  \begin{equation}\label{coneqalseonecomp}
			\operatorname{lim} \operatorname{sup}\underset{S_{c}}{\operatorname{sup}} \frac{1}{N}\left[ I\left(\beta \right)- I\left(\beta^{0} \right)\right] <0 ~ a.s.,
		\end{equation}
		then $\tilde{\beta} \xrightarrow{a.s.} \beta^{0}$ as $N\rightarrow \infty$. 
	\end{lemma}
	\textit{Proof:}
		Proof of this lemma can be obtained in the similar manner to lemma 3 of \cite{alsechirp}.

	\begin{lemma}\label{consleampalseonecomp}
		Suppose $\tilde{\beta}$ is the ALSE of $\beta^{0}$.	Under assumptions 1 and 3, $N^{2}\left(\tilde{\beta}-\beta^{0}\right) \xrightarrow{a.s.} 0$ as $N \rightarrow \infty$.
	\end{lemma}
	\textit{Proof:}
		This result can be proved in the similar manner to lemma 8 of \cite{alsechirplike}.

	\noindent \textit{Proof of Theorem \ref{consthonecompalse}}: Consider the following difference, to derive the consistency of ALSE $\tilde{\beta}$ :
	\begin{equation*}
		\frac{1}{N}\left[ I\left(\beta \right)- I\left(\beta^{0} \right)\right].
	\end{equation*}
	Using lemmas 2 and 4 of \cite{lsechirplike}, for some $c>0$, following holds
	\begin{equation*}
		\operatorname{lim} \operatorname{sup}\underset{S_{c}}{\operatorname{sup}} \frac{1}{N}\left[ I\left(\beta \right)- I\left(\beta^{0} \right)\right] <0 ~ a.s. ~ \text{as} ~ N \rightarrow \infty.
	\end{equation*}
	Hence, using lemma \ref{conslealseonecomp}, $\tilde{\beta} \xrightarrow{a.s.} \beta^{0}~\text{as}~N \rightarrow \infty$. \\
	We now derive the consistency of the estimators $\tilde{A}_{R}$ and $\tilde{A}_{I}$.
	\begin{equation}
		\tilde{A}_{R} = \frac{1}{N} \sum_{t=1}^{N} \left(y_{R} \cos\left( \tilde{\beta}t^{2}\right)+ y_{I} \sin\left( \tilde{\beta}t^{2}\right) \right) . 
	\end{equation}
	Expanding $\cos\left( \tilde{\beta}t^{2}\right) $ and  $\sin\left( \tilde{\beta}t^{2}\right) $ around $\beta^{0}$ using Taylor series expansion and using lemmas 2, 4 of \cite{lsechirplike} and lemma \ref{consleampalseonecomp}, we get the consistency of $\tilde{A}_{R}$. Consistency of $\tilde{A}_{I}$ can be obtained in the similar manner.\\
	
	\noindent \textit{Proof of Theorem \ref{asydthonecompalse}}: We can express $Q\left(\bm{\theta} \right) $ as follows:
	\begin{equation*}
		\frac{1}{N}Q\left(\bm{\theta} \right) =\frac{1}{N} \sum_{t=1}^{N} \left| y\left( t\right) -  A e^{i\beta t^{2}}\right|^{2} = C - \frac{1}{N}J\left(\bm{\theta}\right) +o\left(1 \right) .	
	\end{equation*}
	Here, $C=\frac{1}{N}\sum_{t=1}^{N}\left|y\left( t\right) \right|^{2}$ and \\ $\frac{1}{N}J\left(\bm{\theta}\right) =\frac{2}{N}\sum_{t=1}^{N}\left[ y_{R}\left( t\right) \left\lbrace A_{R} \cos\left(\beta t^{2} \right) -A_{I} \sin\left(\beta t^{2} \right) \right\rbrace + y_{I}\left( t\right) \left\lbrace A_{R} \sin\left(\beta t^{2} \right) +A_{I} \cos\left(\beta t^{2} \right) \right\rbrace  \right] - A_{R}^{2}-A_{I}^{2}$.\\
	
	\noindent Now we calculate the first derivative of $  \frac{1}{N}Q\left(\bm{\theta} \right) $ and $\frac{1}{N}J\left(\bm{\theta} \right)$ at $\bm{\theta} =\bm{\theta}^{0} $ and using lemmas 2 and 4 of \cite{lsechirplike} and conjecture 2 of \cite{lsechirplike}, we get the following relation between these functions:
	\begin{equation*}
		\underset{N \rightarrow \infty }{\operatorname{lim}}Q^{'}\left(\bm{\theta}^{0} \right)\bm{D} =-\underset{N \rightarrow \infty }{\operatorname{lim}}J^{'}\left(\bm{\theta}^{0} \right)\bm{D}.
	\end{equation*}
	After substituting $\tilde{A}_{R}$ and $\tilde{A}_{I}$ in $J\left( {\bm{\theta}}\right) $, we obtain: \begin{equation*}
		J\left( \tilde{A}_{R}, \tilde{A}_{I}, \beta \right) = 2 I\left( \beta \right). 
	\end{equation*}
	Thus, the estimator of $ \bm{\theta}^{0} $ which maximises $J\left(\bm{\theta} \right) $ is equivalent to the ALSE of $ \bm{\theta}^{0} $, i.e. $\tilde{\bm{\theta}}$. Expand $J^{'}\left(\tilde{\bm{\theta}} \right) $ using multivariate Taylor series expansion around the point $ \bm{\theta}^{0} $ and obtain
	\begin{equation*}
		\left(\tilde{\bm{\theta}}-\bm{\theta}^{0} \right) = -J^{'}\left( \bm{\theta}^{0}\right) \left[ J^{''}\left(\bar{\bm{\theta}} \right) \right]^{-1} 
	\end{equation*}
	\begin{equation*}
		\Rightarrow	\left(\tilde{\bm{\theta}}-\bm{\theta}^{0} \right)\bm{D}^{-1} = -\left[ J^{'}\left( \bm{\theta}^{0}\right)\bm{D} \right] \left[\bm{D} J^{''}\left(\bar{\bm{\theta}} \right) \bm{D}\right]^{-1} .
	\end{equation*}
	Here, $\bar{\bm{\theta}}$ lies between $\bm{\theta}^{0}$ and $\tilde{\bm{\theta}}$. Thus, we have $\underset{N \rightarrow \infty }{\operatorname{lim}}\bm{D} J^{''}\left(\bar{\bm{\theta}} \right) \bm{D}=\underset{N \rightarrow \infty }{\operatorname{lim}}\bm{D} J^{''}\left(\bm{\theta}^{0} \right) \bm{D}$. Using lemmas 2 and 4 of \cite{lsechirplike}, we obtain the following relation
	\begin{equation*}
		\underset{N \rightarrow \infty }{\operatorname{lim}}\bm{D} J^{''}\left(\bm{\theta}^{0} \right) \bm{D} = -\underset{N \rightarrow \infty }{\operatorname{lim}}\bm{D} Q^{''}\left(\bm{\theta}^{0} \right) \bm{D} = - \Sigma,
	\end{equation*}
	where $\Sigma$ is as defined in \eqref{sigma}. Thus, we have
	\begin{equation*}
	\begin{split}
		&\left(\tilde{\bm{\theta}}-\bm{\theta}^{0} \right)\bm{D}^{-1} = -\left[ J^{'}\left( \bm{\theta}^{0}\right)\bm{D} \right] \left[\bm{D} J^{''}\left(\bar{\bm{\theta}} \right) \bm{D}\right]^{-1}.\\ &	\Rightarrow	 	\underset{N \rightarrow \infty }{\operatorname{lim}}\left(\tilde{\bm{\theta}}-\bm{\theta}^{0} \right)\bm{D}^{-1} = -\underset{N \rightarrow \infty }{\operatorname{lim}}\left[ J^{'}\left( \bm{\theta}^{0}\right)\bm{D} \right]\underset{N \rightarrow \infty }{\operatorname{lim}} \left[\bm{D} J^{''}\left(\bar{\bm{\theta}} \right) \bm{D}\right]^{-1}\\ &	\Rightarrow	 	\underset{N \rightarrow \infty }{\operatorname{lim}}\left(\tilde{\bm{\theta}}-\bm{\theta}^{0} \right)\bm{D}^{-1} = -\underset{N \rightarrow \infty }{\operatorname{lim}}\left[ Q^{'}\left( \bm{\theta}^{0}\right)\bm{D} \right]\underset{N \rightarrow \infty }{\operatorname{lim}} \left[\bm{D} Q^{''}\left(\bar{\bm{\theta}} \right) \bm{D}\right]^{-1}.
	\end{split}
\end{equation*}
	Now, using \eqref{taylorlseonecomp3}, we have
	\begin{equation*}
		\underset{N \rightarrow \infty }{\operatorname{lim}}\left(\tilde{\bm{\theta}}-\bm{\theta}^{0} \right)\bm{D}^{-1}=\underset{N \rightarrow \infty }{\operatorname{lim}}\left(\hat{\bm{\theta}}-\bm{\theta}^{0} \right)\bm{D}^{-1}.
	\end{equation*}
	
	\noindent It implies that $\tilde{\bm{\theta}}$ and $\hat{\bm{\theta}}$ are asymptotically equivalent in distribution. Hence the result follows.

	\section*{Appendix B} \label{multicompapp}
	\subsection*{B.1}\textbf{ Proofs of the theoretical results of the LSEs for multiple-component model} \label{lsemulticompapp}
	
	\noindent \textit{Proof of Theorem \ref{asydthmulticomplse}}: Using multivariate Taylor series expansion, expand $Q^{'}(\hat{\bm{v}}) $ 
	around the point $\bm{v}^{0}$,
	\begin{equation} \label{taylorlsemulticomp}
		Q^{'}(\hat{\bm{v}} )-Q^{'}(\bm{v}^{0} ) = (\hat{\bm{v}}-\bm{v}^{0}) Q^{''}(\bar{\bm{v}} ).
	\end{equation}
	Here, $\bar{\bm{v}}$ lies between $\bm{v}^{0}$ and $\hat{\bm{v}}$. We can write \eqref{taylorlsemulticomp} as:
	\begin{equation} \label{taylorlsemulticomp2}
		(\hat{\bm{v}}-\bm{v}^{0})= -Q^{'}(\bm{v}^{0} )[ Q^{''}(\bar{\bm{v}} )]^{-1}.
	\end{equation}
	Now, multiplying both side of \eqref{taylorlsemulticomp2} by $\bm{\mathcal{D}}^{-1}$, where $\bm{\mathcal{D}}=diag(\underset{p~times}{\underbrace{\bm{D}, \dots, \bm{D}}})$ and $\bm{D}=diag\left(\frac{1}{\sqrt{N}}, \frac{1}{\sqrt{N}}, \frac{1}{N^{2}\sqrt{N}}\right) $, we have
	\begin{equation} \label{taylorlsemulticomp3}
		(\hat{\bm{v}}-\bm{v}^{0})\bm{\mathcal{D}}^{-1}= -Q^{'}(\bm{v}^{0} )\bm{\mathcal{D}}[\bm{\mathcal{D}} Q^{''}(\bar{\bm{v}} )\bm{\mathcal{D}}]^{-1}. 
	\end{equation}
	Further, using Lindeberg Feller CLT and lemma 2 of \cite{lsechirplike}, it can be shown that:
	\begin{equation} \label{firdermulticomplse}
		Q^{'}(\bm{v}^{0} ) \bm{\mathcal{D}} \xrightarrow{d} \mathcal{N}_{3p}\left( 0, \sigma^{2} \mathcal{E}\right);~ \text{where},
	\end{equation}  \begin{equation} \label{multisigma}
		\mathcal{E}=\begin{bmatrix}
			\Sigma_{1} &0 & \dots & 0\\
			0&\Sigma_{2}&\dots &0\\
			\vdots & \vdots &\ddots & \vdots \\
			0&0& 0 & \Sigma_{p}
		\end{bmatrix},
	\end{equation}
	with $
	\Sigma_{k}=\begin{bmatrix}
		2 &0 & -\frac{2}{3}A_{Ik}^{0}\\
		0&2&\frac{2}{3}A_{Rk}^{0}\\
		-\frac{2}{3}A_{Ik}^{0} &\frac{2}{3}A_{Rk}^{0}& \frac{2}{5}\left| A_{k}^{0}\right|^{2} 
	\end{bmatrix}  \forall k =1, \dots, p.$

\noindent	Using the fact that $\hat{\bm{v}}\xrightarrow{a.s.} \bm{v}^{0}$ as $N\rightarrow \infty$. We have
	\begin{equation*}
		\underset{N\rightarrow \infty}{\operatorname{lim}} \bm{\mathcal{D}} Q^{''}(\bar{\bm{v}} )\bm{\mathcal{D}} =	\underset{N\rightarrow \infty}{\operatorname{lim}} \bm{\mathcal{D}} Q^{''}(\bm{v}^{0} )\bm{\mathcal{D}}.
	\end{equation*}
	The following can be proved by using lemmas  2 and 4 of \cite{lsechirplike}, 
	\begin{equation}\label{secdermulticomplse}
		\underset{N\rightarrow \infty}{\operatorname{lim}} \bm{\mathcal{D}} Q^{''}(\bm{v}^{0} )\bm{\mathcal{D}}= \mathcal{E}.
	\end{equation}
	where $\mathcal{E}$ is as defined in \eqref{multisigma}. After combining the equations, \eqref{taylorlsemulticomp3},  \eqref{firdermulticomplse} and \eqref{secdermulticomplse}, the desired result is obtained.
	
	\subsection*{B.2}\textbf{ Proofs of the theoretical results of the Sequential LSEs} \label{seqlseapp}
	\begin{lemma} \label{consleseqlse}
		Let us consider $S_{c} = \left\lbrace \bm{\theta}: \left|\bm{\theta}- \bm{\theta}_{1}^{0} \right| > c; \bm{\theta} \in \bm{\Theta}_{1} \right\rbrace $. If for any $  c>0 $,
		\begin{equation} \label{coneqseqlse}
			\operatorname{lim} \operatorname{inf} \underset{\bm{\theta} \in S_{c}}{\operatorname{inf}} \frac{1}{N} \left[Q_{1}\left(\bm{\theta} \right) - Q_{1}\left( \bm{\theta}_{1}^{0}\right)   \right] >0 ~  a.s.,
		\end{equation}
		then $\breve{\bm{\theta}}_{1} \xrightarrow{a.s.} \bm{\theta}_{1}^{0}$ as $N \rightarrow \infty$.
	\end{lemma}
	\textit{Proof:}
		This lemma can be proved in the similar manner to lemma \ref{consle}.\\

	\noindent	\textit{Proof of Theorem \ref{consthseqlse1}:} Note that
	\begin{equation*}
		\frac{1}{N} \left[Q_{1}\left(\bm{\theta} \right) - Q_{1}\left( \bm{\theta}_{1}^{0}\right)   \right]= f\left( \bm{\theta}\right) +g\left( \bm{\theta}\right) ,
	\end{equation*}
	where
\begin{equation*}
	\begin{split}
		f\left( \bm{\theta}\right)=&\frac{1}{N} \sum_{t=1}^{N} \left\lbrace \left(A_{R1}^{0} \cos\left( \beta_{1}^{0} t^{2}\right) - A_{I1}^{0}\sin\left(\beta_{1}^{0} t^{2}\right) - A_{R} \cos\left( \beta t^{2}\right) + A_{I}\sin\left(\beta t^{2}\right) \right)^{2} \right. \\ & \left.+ \left(A_{I1}^{0} \cos\left( \beta_{1}^{0} t^{2}\right) + A_{R1}^{0}\sin\left(\beta_{1}^{0} t^{2}\right) - A_{I} \cos\left( \beta t^{2}\right) - A_{R}\sin\left(\beta t^{2}\right) \right)^{2}\right\rbrace +\\ &\frac{2}{N}  \sum_{t=1}^{N} \left\lbrace \left(A_{R1}^{0} \cos\left( \beta_{1}^{0} t^{2}\right) - A_{I1}^{0}\sin\left(\beta_{1}^{0} t^{2}\right) - A_{R} \cos\left( \beta t^{2}\right) + A_{I}\sin\left(\beta t^{2}\right) \right) \right. \\ & \left(\sum_{k=2}^{p} A_{Rk}^{0} \cos\left( \beta_{k}^{0} t^{2}\right) - A_{Ik}^{0}\sin\left(\beta_{k}^{0} t^{2}\right)\right) + \\ & \left(A_{I1}^{0} \cos\left( \beta_{1}^{0} t^{2}\right) + A_{R1}^{0}\sin\left(\beta_{1}^{0} t^{2}\right) - A_{I} \cos\left( \beta t^{2}\right) - A_{R}\sin\left(\beta t^{2}\right) \right)\\ & \left.\left(\sum_{k=2}^{p} A_{Rk}^{0} \sin\left( \beta_{k}^{0} t^{2}\right) + A_{Ik}^{0}\cos\left(\beta_{k}^{0} t^{2}\right)\right) \right\rbrace ,
	\end{split}
\end{equation*}
and 
\begin{equation*}
	\begin{split}
		g\left( \bm{\theta}\right)=&\frac{2}{N} \sum_{t=1}^{N} \left\lbrace \epsilon_{R}\left(t \right) \left(A_{R1}^{0} \cos\left( \beta_{1}^{0} t^{2}\right) - A_{I1}^{0}\sin\left(\beta_{1}^{0} t^{2}\right) - A_{R} \cos\left( \beta t^{2}\right) + A_{I}\sin\left(\beta t^{2}\right) \right) \right. \\ & \left.+\epsilon_{I}\left(t \right) \left(A_{I1}^{0} \cos\left( \beta_{1}^{0} t^{2}\right) + A_{R1}^{0}\sin\left(\beta_{1}^{0} t^{2}\right) - A_{I} \cos\left( \beta t^{2}\right) - A_{R}\sin\left(\beta t^{2}\right) \right)\right\rbrace .
	\end{split}
\end{equation*}
	
	Using lemma 4 of \cite{lsechirplike}, \(	\underset{N\rightarrow \infty}{\operatorname{lim}}\underset{\bm{\theta}\in S_{c}}{\operatorname{sup}} g\left(\bm{\theta} \right)  = 0 ~ a.s.\). Hence, \begin{equation*}
		\operatorname{lim} \operatorname{inf} \underset{\bm{\theta} \in S_{c}}{\operatorname{inf}} \frac{1}{N} \left[Q_{1}\left(\bm{\theta} \right) - Q_{1}\left( \bm{\theta}_{1}^{0}\right)   \right] = \operatorname{lim} \operatorname{inf} \underset{\bm{\theta} \in S_{c}}{\operatorname{inf}}f\left( \bm{\theta}\right) .
	\end{equation*}
	For simplicity, let us assume $p=2$. Consider $S_{c} = \left\lbrace \bm{\theta}: \left|\bm{\theta}-\bm{\theta}_{1}^{0} \right| \geq 3c; \bm{\theta} \in \bm{\Theta}_{1} \right\rbrace \subset S_{c1} \cup S_{c2} \cup S_{c3}  = S$, where,
	\begin{equation*}
	\begin{split}
		S_{c1}&=\left\lbrace \bm{\theta}: \left| A_{R} - A_{R1}^{0} \right| \geq c; \bm{\theta} \in \bm{\Theta}  \right\rbrace \subset \left\lbrace \bm{\theta}: \left| A_{R} - A_{R1}^{0} \right| \geq c; \bm{\theta} \in \bm{\Theta}, \beta = \beta_{1}^{0}  \right\rbrace \\ & \cup \left\lbrace \bm{\theta}: \left| A_{R} - A_{R1}^{0} \right| \geq c; \bm{\theta} \in \bm{\Theta}, \beta= \beta_{2}^{0}, \left(A_{R}, A_{I} \right) = \left(A_{R2}^{0}, A_{I2}^{0} \right)  \right\rbrace \\ &
		\cup \left\lbrace \bm{\theta}: \left| A_{R} - A_{R1}^{0} \right| \geq c; \bm{\theta} \in \bm{\Theta}, \beta= \beta_{2}^{0}, \left(A_{R}, A_{I} \right) \neq \left(A_{R2}^{0}, A_{I2}^{0} \right)  \right\rbrace \\ & 	\cup \left\lbrace \bm{\theta}: \left| A_{R} - A_{R1}^{0} \right| \geq c; \bm{\theta} \in \bm{\Theta}, \beta \neq \beta_{k}^{0}, k=1,2  \right\rbrace ,
	\end{split}
\end{equation*}
\begin{equation*}
	\begin{split}
		S_{c2}&=\left\lbrace \bm{\theta}: \left| A_{I} - A_{I1}^{0} \right| \geq c; \bm{\theta} \in \bm{\Theta}  \right\rbrace \subset \left\lbrace \bm{\theta}: \left| A_{I} - A_{I1}^{0} \right| \geq c; \bm{\theta} \in \bm{\Theta}, \beta = \beta_{1}^{0}  \right\rbrace \\ & \cup \left\lbrace \bm{\theta}: \left| A_{I} - A_{I1}^{0} \right| \geq c; \bm{\theta} \in \bm{\Theta}, \beta= \beta_{2}^{0}, \left(A_{R}, A_{I} \right) = \left(A_{R2}^{0}, A_{I2}^{0} \right)  \right\rbrace \\ &
		\cup \left\lbrace \bm{\theta}: \left| A_{I} - A_{I1}^{0} \right| \geq c; \bm{\theta} \in \bm{\Theta}, \beta= \beta_{2}^{0}, \left(A_{R}, A_{I} \right) \neq \left(A_{R2}^{0}, A_{I2}^{0} \right)  \right\rbrace \\ & 	\cup \left\lbrace \bm{\theta}: \left| A_{I} - A_{I1}^{0} \right| \geq c; \bm{\theta} \in \bm{\Theta}, \beta \neq \beta_{k}^{0}, k=1,2  \right\rbrace ,
	\end{split}
\end{equation*}
\begin{equation*}
	\begin{split}
		S_{c3}&=\left\lbrace \bm{\theta}: \left| \beta - \beta_{1}^{0} \right| \geq c; \bm{\theta} \in \bm{\Theta}  \right\rbrace \subset  \left\lbrace \bm{\theta}: \left| \beta - \beta_{1}^{0} \right| \geq c; \bm{\theta} \in \bm{\Theta}, \beta= \beta_{2}^{0}, \left(A_{R}, A_{I} \right) = \left(A_{R2}^{0}, A_{I2}^{0} \right)  \right\rbrace \\ &
		\cup \left\lbrace \bm{\theta}: \left| \beta - \beta_{1}^{0} \right| \geq c; \bm{\theta} \in \bm{\Theta}, \beta= \beta_{2}^{0}, \left(A_{R}, A_{I} \right) \neq \left(A_{R2}^{0}, A_{I2}^{0} \right)  \right\rbrace \\ & 	\cup \left\lbrace \bm{\theta}: \left| \beta - \beta_{1}^{0} \right| \geq c; \bm{\theta} \in \bm{\Theta}, \beta \neq \beta_{k}^{0}, k=1,2  \right\rbrace .
	\end{split}
\end{equation*}
	Using lemma 2 of \cite{lsechirplike} for each of the above set, we have that $\operatorname{lim} \operatorname{inf} \underset{\bm{\theta} \in T}{\operatorname{inf}}f\left( \bm{\theta}\right) >0$ a.s., where $T$ can be any of these sets $	S_{c1}, S_{c2}$ or $	S_{c3}$. Therefore, the result follows.

	\begin{lemma} \label{consleseqlse2}
		Under assumptions 1, 4 and 5,  $\left( \breve{\bm{\theta}}_{1}- \bm{\theta}_{1}^{0}\right)\left( \sqrt{N} \bm{D}\right)^{-1}  \xrightarrow{a.s.} 0 $ as $N \rightarrow \infty$,	where $\bm{D}=diag\left(\frac{1}{\sqrt{N}}, \frac{1}{\sqrt{N}}, \frac{1}{N^{2}\sqrt{N}}\right) $.
	\end{lemma}
	\textit{Proof:} This lemma can be proved in the similar manner of lemma 4 of \cite{seqlsechirp}.\\

	\noindent	\textit{Proof of Theorem \ref{consthseqlse2}}: Using the above lemma \ref{consleseqlse2}, we have:\begin{equation*}
		\breve{A}_{R1}= A_{R1}^{0}+o\left(1 \right),~\breve{A}_{I1}= A_{I1}^{0}+o\left(1 \right)~ \text{and}~	\breve{\beta_{1} }= \beta_{1}^{0}+o\left(N^{-2} \right).
	\end{equation*}
Thus, we have,
	\begin{equation} \label{seqlseupd}
		\breve{A}_{1}e^{i\breve{\beta}_{1}t^{2}}=A_{1}^{0}e^{i \beta_{1}^{0}t^{2}}+o\left( 1\right) .
	\end{equation}
	Using \eqref{seqlseupd} and following the similar arguments as in theorem 7, the result follows.
	
	\noindent	The proof of the theorem \ref{consthseqlse3} directly follows from the lemma stated below.
	
	\begin{lemma}\label{consleseqlse3}
		If $\epsilon\left( t\right) $ satisfies the assumption 1, and $\breve{A}_{R}, \breve{A}_{I}$ and $\breve{\beta}$ are obtained by minimizing the function given as follows:\begin{equation*}
			Q_{\left( p+1\right) }\left( \bm{\theta}\right) =\frac{1}{N}\sum_{t=1}^{N}\left|\epsilon\left( t\right)-Ae^{i \beta t^{2}}  \right|^{2},
		\end{equation*}
		then $\breve{A}_{R} \xrightarrow{a.s.}0$ and $\breve{A}_{I} \xrightarrow{a.s.}0$.
	\end{lemma}
	\textit{Proof:}
		Note that
		\begin{equation*}
			Q_{\left( p+1\right) }\left( \bm{\theta}\right)=R\left(\bm{\theta} \right) +o\left( 1\right), ~\text{where},
		\end{equation*}
		\begin{equation*}
			\begin{split}
				R\left( \bm{\theta}\right) =&\frac{1}{N} \sum_{t=1}^{N}\left|\epsilon\left( t\right)  \right|^{2} -\frac{2}{N}\sum_{t=1}^{N}\left\lbrace \epsilon_{I}\left(t \right)\left(  A_{I}\cos\left(\beta t^{2} \right) + A_{R}\sin\left(\beta t^{2} \right)\right)+\epsilon_{R}\left(t \right) \right.\\&\left.\left(  A_{R}\cos\left(\beta t^{2} \right) - A_{R}\sin\left(\beta t^{2} \right)\right) \right\rbrace 
				+A_{R}^{2}+A_{I}^{2}.
			\end{split}
		\end{equation*}
		Since $R\left(\bm{\theta}\right) $ and $Q_{\left( p+1\right) }\left( \bm{\theta}\right)$ are equivalent as $N \rightarrow \infty$, therefore, estimators of $R\left(\bm{\theta}\right) $ and $Q_{\left( p+1\right) }\left( \bm{\theta}\right)$ would be equivalent. Thus, we have:
		\begin{equation*}
			\breve{A}_{R} =\frac{1}{N} \sum_{t=1}^{N}\epsilon_{R}\left( t\right) \cos\left(\breve{\beta} t^{2}\right) +\epsilon_{I}\left( t\right) \sin\left(\breve{\beta} t^{2}\right)+o\left(1 \right) .
		\end{equation*}
		\begin{equation*}
			\breve{A}_{I} =\frac{1}{N} \sum_{t=1}^{N}\epsilon_{I}\left( t\right) \cos\left(\breve{\beta} t^{2}\right) -\epsilon_{R}\left( t\right) \sin\left(\breve{\beta} t^{2}\right)+o\left(1 \right) .
		\end{equation*}
		Using lemma 4 of \cite{lsechirplike}, we obtain the desired result.\\

\noindent	\textit{Proof of Theorem \ref{asydthseqlse1}}: The error sum of squares $Q_{1}\left( \bm{\theta} \right)$ is expressed as:	\begin{equation}
		Q_{1}\left( \bm{\theta} \right) =\sum_{t=1}^{N} \left| y_{1}\left( t\right) -  A e^{i\beta t^{2}}\right|^{2}.
	\end{equation}
	Expanding $Q_{1}^{'}(\breve{\bm{\theta}}_{1} ) $ using Taylor series expansion, we get:
	\begin{equation} \label{asydtaylorseqlse3}
		(\breve{\bm{\theta}}_{1}-\bm{\theta}_{1}^{0})\bm{D}^{-1}= -Q_{1}^{'}(\bm{\theta}_{1}^{0} )\bm{D}[\bm{D} Q_{1}^{''}(\bar{\bm{\theta}_{1}} )\bm{D}]^{-1}.
	\end{equation}
	Using Lindeberg Feller CLT and lemma 2 of \cite{lsechirplike} and conjecture 2 of \cite{lsechirplike}, it can be shown that:
	
	\begin{equation} \label{asydfirderseqlse}
		Q_{1}^{'}(\bm{\theta}_{1}^{0} ) \bm{D} \xrightarrow{d} \mathcal{N}_{3}\left( 0, \sigma^{2} \Sigma_{1}\right);~ \text{where},
	\end{equation} 	\begin{equation} \label{sigma1}
		\Sigma_{1}=\begin{bmatrix}
			2 &0 & -\frac{2}{3}A_{I1}^{0}\\
			0&2&\frac{2}{3}A_{R1}^{0}\\
			-\frac{2}{3}A_{I1}^{0} &\frac{2}{3}A_{R1}^{0}& \frac{2}{5}\left| A_{1}^{0}\right|^{2} 
		\end{bmatrix}.\end{equation}
	Using $\breve{\bm{\theta}}_{1}\xrightarrow{a.s.} \bm{\theta}_{1}^{0}$ as $N\rightarrow \infty$,
	\begin{equation*}
		\underset{N\rightarrow \infty}{\operatorname{lim}} \bm{D} Q_{1}^{''}(\bar{\bm{\theta}}_{1} )\bm{D} =	\underset{N\rightarrow \infty}{\operatorname{lim}} \bm{D} Q_{1}^{''}(\bm{\theta}_{1}^{0} )\bm{D}.
	\end{equation*}
	\noindent Using lemmas 2 and 4 of \cite{lsechirplike}, it can be shown that:	\begin{equation}\label{asydsecderseqlse}
		\underset{N\rightarrow \infty}{\operatorname{lim}} \bm{D} Q_{1}^{''}(\bm{\theta}_{1}^{0} )\bm{D}= \Sigma_{1}.
	\end{equation}
	Using \eqref{asydtaylorseqlse3},  \eqref{asydfirderseqlse} and \eqref{asydsecderseqlse}, we get the asymptotic distribution of the estimators of the first component. Asymptotic distribution of $\breve{\bm{\theta}}_{k}$ for  $ k=2$ can be obtained along the similar lines using lemma \ref{consleseqlse2}. Thus, this result can be proved for all $ k=3, \dots, p$ by using the similar arguments.
	
	\subsection*{B.3}\textbf{ Proofs of the theoretical properties of the Sequential ALSEs}\label{seqalseapp}
	\begin{lemma} \label{consleseqalse}
		Consider the following set $S_{c}=\left\lbrace \beta: \left| \beta - \beta_{1}^{0} \right|>c \right\rbrace $. If for any $c>0$, \begin{equation}\label{coneqalsmulticomp}
			\operatorname{lim} \operatorname{sup}\underset{S_{c}}{\operatorname{sup}} \frac{1}{N}\left[ I_{1}\left(\beta \right)- I_{1}\left(\beta_{1}^{0} \right)\right] <0 ~ a.s.,
		\end{equation}
		then $\tilde{\beta}_{1} \xrightarrow{a.s.} \beta_{1}^{0}$ as $N\rightarrow \infty$.
	\end{lemma}
	\textit{Proof:}
		This proof can be obtained by using the similar arguments as in the lemma \ref{conslealseonecomp}.

	\begin{lemma}\label{consleseqalse2}
		Suppose $\tilde{\beta}_{1}$ is the ALSE of $\beta_{1}^{0}$. If assumptions 1, 6 and 7 are satisfied, then $N^{2}\left(\tilde{\beta}_{1}-\beta_{1}^{0}\right) \xrightarrow{a.s.} 0$ as $N \rightarrow \infty$.
	\end{lemma}
	\textit{Proof:} This lemma can be proved using the similar arguments as in the lemma \ref{consleampalseonecomp}.\\

	\noindent	\textit{Proof of Theorem \ref{consthseqalse1}}:  We first derive strong consistency of $\tilde{\beta}_{1}$. For simplicity, assume that $ p=2 $. \\
	The set $S_{c}=\left\lbrace \beta: \left| \beta - \beta_{1}^{0} \right|>c \right\rbrace $ can be split into two parts as follows: $S_{c}\subset S_{c}^{1} \cup S_{c}^{2}$; where $S_{c}^{1}=\left\lbrace \beta: \left| \beta - \beta_{1}^{0} \right|>c; \beta=\beta_{2}^{0} \right\rbrace$ and $S_{c}^{2}=\left\lbrace \beta: \left| \beta - \beta_{1}^{0} \right|>c; \beta\neq\beta_{2}^{0} \right\rbrace$.
	Using lemmas 2 and 4 of \cite{lsechirplike}, it can be shown that for some $c>0$;
	\begin{equation*}
		\operatorname{lim} \operatorname{sup}\underset{S_{c}^{1}}{\operatorname{sup}} \frac{1}{N}\left[ I_{1}\left(\beta \right)- I_{1}\left(\beta_{1}^{0} \right)\right] <0 ~ a.s. ~ \text{as} ~ N \rightarrow \infty.	
	\end{equation*}	\begin{equation*}
		\operatorname{lim} \operatorname{sup}\underset{S_{c}^{2}}{\operatorname{sup}} \frac{1}{N}\left[ I_{1}\left(\beta \right)- I_{1}\left(\beta_{1}^{0} \right)\right]<0 ~a.s. ~ \text{as} ~  N \rightarrow \infty.
	\end{equation*}
	Hence, using lemma \ref{consleseqalse}, $\tilde{\beta}_{1} \xrightarrow{a.s.} \beta_{1}^{0}~ \text{as}~ N \rightarrow \infty$. \\
	Strong consistency of the estimators $\tilde{A}_{R1}$ and $\tilde{A}_{I1}$ can be proved using the similar arguments used as in theorem \ref{consthonecompalse}.\\
	
	\noindent	\textit{Proof of Theorem \ref{consthseqalse2}}: Using lemma \ref{consleseqalse2} and theorem \ref{consthseqalse1}, \begin{equation*}
		\tilde{A}_{R1}= A_{R1}^{0}+o\left(1 \right) ,~	\tilde{A}_{I1}= A_{I1}^{0}+o\left(1 \right)  ~\text{and}~	\tilde{\beta_{1} }= \beta_{1}^{0}+o\left(N^{-2} \right).
	\end{equation*}
	Thus,
	\begin{equation} \label{seqalseupd}
		\tilde{A}_{1}e^{i\tilde{\beta}_{1}t^{2}}=A_{1}^{0}e^{i \beta_{1}^{0}t^{2}}+o\left( 1\right). 
	\end{equation}
	Now, using \eqref{seqalseupd} and following the similar arguments as in theorem \ref{consthseqalse1}, we obtain the desired result.	\\
	
\noindent	\textit{Proof of Theorem \ref{consthseqalse4}}: Note that $\tilde{A}_{R\left(p+1 \right) }$ and $\tilde{A}_{I\left(p+1 \right) }$ are the ALSEs of the $A_{R\left(p+1 \right) }^{0}$ and $A_{I\left(p+1 \right) }^{0}$, respectively,
	\begin{equation*}
		\tilde{A}_{R\left(p+1 \right) } =\frac{1}{N} \sum_{t=1}^{N} \left(y_{R\left(p+1 \right)} \cos\left( \tilde{\beta}_{p+1}t^{2}\right)+ y_{I\left(p+1 \right)} \sin\left( \tilde{\beta}_{p+1}t^{2}\right) \right) ,
	\end{equation*}
	\begin{equation*}
		\tilde{A}_{I\left(p+1 \right) } =\frac{1}{N} \sum_{t=1}^{N} \left(y_{I\left(p+1 \right)} \cos\left( \tilde{\beta}_{p+1}t^{2}\right)- y_{R\left(p+1 \right)} \sin\left( \tilde{\beta}_{p+1}t^{2}\right) \right) ,
	\end{equation*}
	where, $y_{p+1}$ is the data obtained by taking out the effect of first p chirp components from the original data $y_{1}\left(t \right) $, and $y_{R\left(p+1 \right)}$ and $y_{I\left(p+1 \right)}$ are the real and imaginary parts of the data $y_{p+1}$, respectively. It implies that: 
	\begin{equation} \label{updrealcomp}
		y_{R\left(p+1 \right)}=y_{R1}\left( t\right) - \sum_{k=1}^{p}\left( \tilde{A}_{Rk}\cos\left(\tilde{\beta}_{k}t^{2} \right)-\tilde{A}_{Ik}\sin\left(\tilde{\beta}_{k}t^{2} \right) \right),
	\end{equation}
	\begin{equation} \label{updimgcomp}
		y_{I\left(p+1 \right)}=y_{I1}\left( t\right) - \sum_{k=1}^{p}\left( \tilde{A}_{Rk}\sin\left(\tilde{\beta}_{k}t^{2} \right)+\tilde{A}_{Ik}\cos\left(\tilde{\beta}_{k}t^{2} \right) \right).
	\end{equation}
	Using theorem \ref{consthseqalse3}, \eqref{updrealcomp} and \eqref{updimgcomp} can be expressed as: \(	y_{R\left(p+1 \right)}=\epsilon_{R}\left(t \right) +o\left( 1\right) \) and \(y_{I\left(p+1 \right)}=\epsilon_{I}\left(t \right) +o\left( 1\right) \). 
	Using lemma 4 of \cite{lsechirplike}, we obtain: 	$\tilde{A}_{R\left(p+1 \right) } \xrightarrow{a.s.} 0$, $\tilde{A}_{I\left(p+1 \right) } \xrightarrow{a.s.} 0$ as $N \rightarrow \infty$. Hence, the result follows.\\
	
	\noindent	\textit{Proof of Theorem \ref{asydthseqalse1}}: We first derive that the asymptotic distribution of $( \tilde{\bm{\theta}}_{1} - \bm{\theta}_{1}^{0}) \bm{D}^{-1}$ is identical as that of the $( \breve{\bm{\theta}}_{1} - \bm{\theta}_{1}^{0}) \bm{D}^{-1}$.
	We have
	\begin{equation*}
		\frac{1}{N}Q_{1}\left(\bm{\theta} \right) = C_{1} - \frac{1}{N}J_{1}\left(\bm{\theta}\right) +o\left(1 \right) .	
	\end{equation*}	Here, \(C_{1}=\frac{1}{N}\sum_{t=1}^{N}\left|y_{1}\left( t\right) \right|^{2}\) and\\ $\frac{1}{N}J_{1}\left(\bm{\theta}\right) =\frac{2}{N}\sum_{t=1}^{N}\left[ y_{R1}\left( t\right) \left\lbrace A_{R} \cos\left(\beta t^{2} \right) -A_{I} \sin\left(\beta t^{2} \right) \right\rbrace + y_{I1}\left( t\right) \left\lbrace A_{R} \sin\left(\beta t^{2} \right) +A_{I} \cos\left(\beta t^{2} \right) \right\rbrace  \right] - A_{R}^{2}-A_{I}^{2}$.\\	Now proceeding on the similar lines as in theorem 4, we can show that the asymptotic distributions of $\tilde{\bm{\theta}}_{1}$ and $\breve{\bm{\theta}}_{1}$ are identical.\\
	Asymptotic distribution of $\tilde{\bm{\theta}}_{k}$ for $ k=2$ can be obtained along the similar lines using lemma \ref{consleseqalse2}. Hence this result can be proved for all $ k=3, \dots, p$ by using the similar arguments.

\end{document}